\documentclass[twocolumn]{aastex63}

\newcommand{\hp}{\mbox{HAT-P-23}}
\newcommand{\hpb}{\mbox{HAT-P-23b}}

\newcommand{\angstrom}{\textup{\AA}}
\newcommand{\halpha}{$\rm H\alpha$}
\newcommand{\water}{$\rm H_2O$\xspace}
\newcommand{\Na}{\ion{Na}{1}\xspace}
\newcommand{\K}{\ion{K}{1}\xspace}
\newcommand{\Msun}{M_\odot}
\newcommand{\Rsun}{R_\odot}

\newcommand{\Ms}{M_\text{s}}
\newcommand{\Rs}{R_\text{s}}
\newcommand{\Mj}{M_\text{J}}
\newcommand{\Rj}{R_\text{J}}
\newcommand{\Mp}{M_\text{p}}
\newcommand{\Rp}{R_\text{p}}

\newcommand{\RpRs}{\Rp/\Rs}

\newcommand{\logg}{\log\text{ g (cm/s$^2$)}}
\newcommand{\Teff}{T_\text{eff}}
\newcommand{\Teq}{T_\text{eq}}
\newcommand{\Tp}{T_\text{p}}
\newcommand{\Tchord}{T_\text{star}}
\newcommand{\Thet}{T_\text{het}}
\newcommand{\fhet}{f_\text{het}}

\newcommand\pylink[1]{\href{#1}{\texttt{(</>)}}}

\newcommand{\PlanetSignal}{$\SI{384}{ppm}$\xspace}
\newcommand{\MeanWLCDepth}{$\SI{12966}{ppm}$\xspace}
\newcommand{\MeanWLCDepthErr}{$\SI{12966 \pm 228}{ppm}$\xspace}
\newcommand{\MeanWLCDepthSp}{$\SI{12948}{ppm}$\xspace}
\newcommand{\MeanWLCDepthErrSp}{$\SI{12948 \pm 241}{ppm}$\xspace}

\usepackage{graphicx, amsmath, bm, hyperref, ulem, xspace}
\usepackage[hyphens]{url}
\hypersetup{linkcolor=red,citecolor=blue,filecolor=cyan,urlcolor=cyan}
\usepackage{savesym}
\savesymbol{tablenum}
\usepackage[detect-weight=true, detect-family=true]{siunitx}
\restoresymbol{SIX}{tablenum}
\usepackage[T1]{fontenc}
\usepackage{hyphenat}
\usepackage[english]{babel}
\useshorthands{"}
\defineshorthand{"-}{\babelhyphen{hard}}
\usepackage[final]{microtype}
\usepackage[hang,flushmargin]{footmisc}
\received{2021 February 12}
\revised{2021 April 05}
\accepted{2021 April 07}
\shorttitle{ACCESS: \hpb{}}
\shortauthors{Weaver et al.}
\graphicspath{{./}{figures/}}
\begin{document}
\title{ACCESS: An optical transmission spectrum of the high"-gravity, hot Jupiter \hpb{}}

\correspondingauthor{Ian C. Weaver}
\email{iweaver@cfa.harvard.edu}

\author[0000-0001-6205-6315]{Ian C. Weaver}
\affiliation{Center for Astrophysics ${\rm \mid}$ Harvard {\rm \&} Smithsonian,
60 Garden St, Cambridge, MA 02138, USA}

\author[0000-0003-3204-8183]{Mercedes L\'opez-Morales}
\affiliation{Center for Astrophysics ${\rm \mid}$ Harvard {\rm \&} Smithsonian,
60 Garden St, Cambridge, MA 02138, USA}

\author[0000-0003-4157-832X]{Munazza K. Alam}
\affiliation{Center for Astrophysics ${\rm \mid}$ Harvard {\rm \&} Smithsonian,
60 Garden St, Cambridge, MA 02138, USA}

\author[0000-0001-9513-1449]{N\'estor Espinoza}
\affiliation{Space Telescope Science Institute, 3700 San Martin Drive, Baltimore, MD 21218, USA}

\author[0000-0002-3627-1676]{Benjamin V. Rackham}
\altaffiliation{51 Pegasi b Fellow}
\affiliation{Department of Earth, Atmospheric and Planetary Sciences, and
Kavli Institute for Astrophysics and Space Research, Massachusetts Institute
of Technology, Cambridge, MA 02139, USA}

\author[0000-0002-8515-7204]{Jayesh M. Goyal}
\affiliation{Department of Astronomy and Carl Sagan Institute,
Cornell University, 122 Sciences Drive, 14853, Ithaca, NY, USA}

\author[0000-0003-4816-3469]{Ryan J. MacDonald}
\affiliation{Department of Astronomy and Carl Sagan Institute,
Cornell University, 122 Sciences Drive, 14853, Ithaca, NY, USA}

\author[0000-0002-8507-1304]{Nikole K. Lewis}
\affiliation{Department of Astronomy and Carl Sagan Institute,
Cornell University, 122 Sciences Drive, 14853, Ithaca, NY, USA}

\author[0000-0003-3714-5855]{D\'aniel Apai}
\affiliation{Department of Astronomy/Steward Observatory, The
University of Arizona, 933 N. Cherry Avenue, Tucson, AZ 85721, USA}
\affiliation{Earths in Other Solar Systems Team, NASA Nexus
for Exoplanet System Science}
\affiliation{Lunar and Planetary Laboratory, The University of Arizona,
1640 E. Univ. Blvd, Tucson, AZ 85721}

\author[0000-0003-2831-1890]{Alex Bixel}
\affiliation{Department of Astronomy/Steward e\.g\.,vatory, The
University of Arizona, 933 N. Cherry Avenue, Tucson, AZ 85721, USA}
\affiliation{Earths in Other Solar Systems Team, NASA Nexus for Exoplanet System
Science}

\author[0000-0002-5389-3944]{Andr\'es Jord\'an}
\affiliation{Facultad de Ingenier\'ia y Ciencias, Universidad Adolfo
Ib\'a\~nez, Av.\ Diagonal las Torres 2640, Pe\~nalol\'en, Santiago, Chile}
\affiliation{Millennium Institute for Astrophysics,  Santiago, Chile}

\author[0000-0002-4207-6615]{James Kirk}
\affiliation{Center for Astrophysics ${\rm \mid}$ Harvard {\rm \&} Smithsonian,
60 Garden St, Cambridge, MA 02138, USA}

\author[0000-0002-6167-3159]{Chima McGruder}
\affiliation{Center for Astrophysics ${\rm \mid}$ Harvard {\rm \&} Smithsonian,
60 Garden St, Cambridge, MA 02138, USA}

\author{David J. Osip}
\affiliation{Las Campanas Observatory, Carnegie Institution for Science, Colina el Pino, Casilla 601 La Serena, Chile}

%\author{Jennifer Fienco}
%\affiliation{Instituto de Astrof\'isica, Facultad de
%F\'isica, Pontificia Universidad Cat\'olica de Chile, Av. Vicu\~na Mackenna
%4860, 7820436 Macul, Santiago, Chile}



\begin{abstract}
We present a new ground"-based visible transmission spectrum of the high-gravity, hot
Jupiter \hpb{}, obtained as part of the ACCESS project. We derive the spectrum from five
transits observed between 2016 and 2018, with combined wavelength coverage between
\SIrange{5200}{9269}{\angstrom} in \SI{200}{\angstrom} bins, and with a median precision
of \SI{247}{ppm} per bin. HAT-P-23b's relatively high surface gravity
$(g \approx \SI{30}{m/s^2})$, combined with updated stellar and planetary parameters from
\textit{Gaia} DR2, gives a 5-scale-height signal of \PlanetSignal for a hydrogen-dominated
atmosphere. Bayesian models favor a clear atmosphere for the planet with the tentative
presence of TiO, after simultaneously modeling stellar contamination, using spots
parameter constraints from photometry. If confirmed, HAT-P-23b would be the first example
of a high-gravity gas giant with a clear atmosphere observed in transmission at optical/NIR wavelengths; therefore, we recommend expanding
observations to the UV and IR to confirm our results and further characterize this
planet. This result demonstrates how combining transmission spectroscopy of exoplanet
atmospheres with long-term photometric monitoring of the host stars can help disentangle
the exoplanet and stellar activity signals.
\end{abstract}


%% e\.g\.,ve the use of the LaTeX \label
%% and \citet commands to identify citations. The citations are

\section{Introduction} \label{sec:intro}
Observations of exoplanetary atmospheres offer the possibility of understanding the
atmospheric physical properties and chemical composition of those worlds, as well as
providing clues to their formation and evolution histories (e.g., \citealt{oberg13};
\citealt{moses13}; \citealt{mordasini16}, \citealt{espinoza2017}). The first comparative
studies of exoplanetary atmospheres using transmission spectra (see, e.g.,
\citealt{sing2016}) found evidence of a gradual transition between clear and cloudy
atmospheres, but no clear correlation of that transition with other system parameters,
such as planetary mass, gravity, effective temperature, or stellar irradiation levels, and
the chemical composition of the star.

High-altitude clouds and hazes have been inferred in the atmosphere of a number of
exoplanets from scattering slopes in the visible \citep[e.g.,][]{pont2008, pont2013,
sing2009, sing2011, sing2013, jordan2013, gibson2013}, and from  the absence of the
pressure"-broadened alkali \Na and \K lines originating from deeper within the atmosphere
\citep[e.g.,][]{charbonneau2002,wakeford2014}. The first detections of gas giants with
potentially clear atmospheres, as indicated by the presence of complete
pressure-broadened alkali line profiles, have only recently been made, e.g., WASP"-39b
\citep{nikolov2016, fischer2016, wakeford2018, kirk2019}, WASP"-96b \citep{nikolov2018},
and WASP"-62b \citep{alam2021}.

A current driving focus of this field is identifying what system parameters may correlate
with the observed atmospheric properties of exoplanets. For example, the relationship
between chemical abundance in an exoplanet's atmosphere and planet mass is actively being
explored \citep[e.g.][]{helling2016, sing2016, kreidberg2014, fraine2014}. Recently,
\citet{Pinhas2019} reanalyzed the \citet{sing2016} sample and concluded that the majority
of hot Jupiters have atmospheres consistent with sub"-solar \water abundances. In
addition, comparisons of \water absorption vs. equilibrium temperature
\citep{guangwei2017, stevenson2016}, and cloudiness vs. irradiation and cloudiness vs.
\Na/\K absorption \citep{heng2016} have also been explored. The current population of
exoplanets studied is still too limited to make any substantial claims regarding
statistical correlations between the parameters explored in these studies, making
continued work in exoplanet characterization extremely important for the future of
comparative exoplanetology. Ground"-based (e.g., ACCESS\footnote{ACCESS, previously known as the Arizona"-CfA"-Cat\'olica"-Carnegie Exoplanet
Spectroscopy Survey \citep{rackham2017}, has been renamed to the Atmospheric Characterization Collaboration for Exoplanet Spectroscopic Studies}, GPIES\footnote{Gemini Planet Imager Extra Solar
Survey \citep{nielsen2019}}, VLT FORS2\footnote{Very Large Telescope FOcal Reducer and
Spectrograph \citep{nikolov2018}}, LRG"-BEASTS\footnote{Low Resolution Ground"-Based
Exoplanet Atmosphere Survey using Transmission Spectroscopy \citep{kirk2018}}) and
space"-based surveys (PanCET\footnote{Panchromatic Comparative Exoplanetology Treasury
\citep{wakeford2017}}) are working towards providing homogeneous spectra of statistically
significant numbers of exoplanet atmospheres in the search for these parameter correlations.

For example, the population of high"-gravity exoplanets offers a novel testing ground for planet interior and atmosphere studies. For transmission spectroscopy observations, we often rely on estimates of
the scale height of a planet's atmosphere to gauge the expected amplitude of the
spectroscopic features. A planet's atmospheric scale height ($H$) is fundamentally
tied to its surface gravity ($g$), temperature ($T$), and mean molecular weight ($\mu$), by $H=k_B T/\mu g$. To date, there has been a strong preference to target exoplanets with large scale heights for transmission
spectroscopy observations, which has resulted in the bulk of observational data that is focused
on objects that are much lower gravity and much hotter than Jupiter
($g \sim \SI{25}{m/s^2}$, $T_\text{eff}\sim\SI{150}{K}$)
in our own Solar System \citep[see for example review by][]{sing2018}. Thus
our view of the physics and chemistry at work in exoplanet atmospheres has been obscured
by this biased sample. In particular, a planet's gravity plays a key role in shaping its
horizontal and vertical advection patterns \citep[e.g.,][]{showman2010} and the vertical
extent of clouds in the atmosphere \citep[e.g.,][]{marley2013}. Although technically
challenging, the extension of the sample of exoplanets with transmission spectroscopy
observations into the ``high-gravity'' regime will allow us to tease apart the physical
processes shaping the atmospheres of these distant worlds. To date, the only exoplanets with surface gravities in excess of $\SI{20}{m/s^2}$ to be probed via
transmission spectroscopy at optical-NIR wavelengths are HD~189733b ($g \sim \SI{23}{m/s^2}$, $\Teq \sim \SI{1200}{K}$), WASP-43b ($g \sim \SI{47}{m/s^2}$, $\Teq \sim \SI{1440}{K}$), and now HAT-P-23b $(g \sim \SI{31}{m/s^2},\ \Teq \sim \SI{2000}{K})$ with this
work. In this paper, we present the ground"-based visible transmission spectrum of the hot Jupiter
\hpb{}, obtained as part of the ACCESS survey.

\hpb{} is a hot Jupiter discovered by \citet{bakos2010} transiting the G0
star \hp{}
\citep[$\Ms = \SI{1.08 \pm 0.14}{\Msun}$,
       $\Rs = \SI{1.152 \pm 0.06}{\Rsun}$,
       $\Teff = \SI{5920.0 \pm 140.0}{K}$,
       $V   = 11.9$;][]{stassun2019}.
Assuming a circular orbit ($e<0.052$; \citealt{bonomo2017}), and combining the most recent
transit
\citep[$\RpRs = 0.1113 \pm 0.001$,
       $P = \SI{1.2128867 \pm 0.0000002}{days}$;][]{sada_and_ramon2016}
and radial velocity observations
\citep[$K_\text{RV} = \SI{346 \pm 21}{m/s}$,
       $i = \SI{85.1 \pm 1.5}{^\circ}$;][]{bonomo2017}
with the updated stellar parameters from \textit{Gaia} parallax measurements,
\citet{stassun2019} give self"-consistent
estimates\footnote{\url{https://icweaver.github.io/ExoCalc.jl/}} for the physical
parameters of \hpb{} as
$
    \Mp = \SI{1.92 \pm 0.20}{\Mj},
    \Rp = \SI{1.247 \pm 0.066}{\Rj},
    \Teq = \SI{2027 \pm 92}{K},\text{ and }
    \logg = 3.485 \pm 0.064
$.

Previous multiband, photometric observations in the visible portion of the spectrum
collected by the Bonn University Simultaneous Camera (BUSCA), mounted on the \SI{2.2}{m}
Calar Alto Telescope, indicate a flat transmission spectrum \citep{ciceri2015}.
Additionally, secondary eclipse data collected in $H$ and $K_s$ bands with the Wide"-field
Infrared Camera (WIRC) on the \SI{200}{\text{inch}} Hale Telescope and in the
\SI{3.6}{\micron} and \SI{4.5}{\micron} bands with the InfraRed Array Camera (IRAC) aboard
Spitzer indicate a planetary atmosphere with low efficiency energy transport from its
day"-side to night"-side, no thermal inversion, and a lack of strong absorption features
\citep{orourke2014}.

We use new transit observations on \textit{Magellan/IMACS} to produce the visible spectrum
spanning \SIrange{0.52}{0.93}{\micron}. We use this transmission spectrum to better
constrain the abundance of \Na and \K, and search for the presence of clouds/hazes in the
atmosphere in \hpb{}. In our analysis, we find that a clear planetary atmosphere
containing TiO is marginally preferred over other models including stellar activity.

This paper is structured as follows. In Section~\ref{sec:obs} we present our
\textit{Magellan/IMACS} observations. In Section~\ref{sec:reduction} we outline the data
reduction process used in our observations. In Section~\ref{sec:detrending} we discuss our
detrending techniques and present the detrended white"-light and binned light curves for
each night. In Section~\ref{sec:stell_act} we investigate the impact of stellar activity.
In Section~\ref{sec:tspec} we present the final combined transmission spectrum, and in
Section~\ref{sec:atmospheric_modeling} we present the results of a forward modeling and
retrieval analysis on this spectrum when the presence of a heterogeneous stellar
photosphere is taken into account. We then summarize and conclude in
Section~\ref{sec:conclusion}. For convenience, we include links throughout the paper to
\texttt{Python} code \pylink{https://icweaver.github.io/HAT-P-23b/README.html}
used to produce each figure, which is archived on Zenodo
\href{https://zenodo.org/record/4673489}{doi:10.5281/zenodo.4673489}.

\section{Observations} \label{sec:obs}
\subsection{General setup} \label{ssec:general_setup}
We observed five transits of \hpb{} between 2016 and 2018 with the 6.5"-meter Magellan
Baade Telescope and Inamori"-Magellan Areal Camera and Spectrograph (IMACS,
\citealt{dressler2006}) as part of ACCESS, a multi"-institutional effort to build a
comparative library of visual transmission spectra of transiting exoplanets. For these
observations, we used the IMACS f/2 camera. This camera has a \SI{27.4}{\arcmin} diameter
field of view (FoV), in which we can observe simultaneously the target and several
comparison stars, allowing for effective removal of common instrumental and atmospheric
systematics. We selected comparison stars less than 0.5 magnitude brighter and 1 magnitude
fainter that \hpb{} and closest in $B-V/J-K$ color space, following \citet{rackham2017}.
We show the selected comparison stars in Table~\ref{tab:comp_stars}.

For ACCESS we have used several combinations of custom multi"-slit masks, grisms, and
blocking filters over time to refine our observations as the survey progresses. For the
first transit of \hpb{}, in particular, we used $20''\times12''$ wide slits for the science masks to
adequately sample the sky background and narrower, $20''\times0.5''$, slits for the
calibration masks to perform high precision wavelength calibration. We used the 150
lines/mm grism blazed at $18.8^\circ$ (150-18.8) with the WB5600 blocking filter for this
setup. We switched to a new field placement and a new science mask with wider
$90''\times10''$ slits for the last four transits to better sample the sky and get more
uniform wavelength coverage. We used the 300 lines/mm grism blazed at $17.5^\circ$
(300-17.5) with the GG495 blocking filter for the second through third transit, the 300
lines/mm grism blazed at $17.5^\circ$ (300-17.5) with the GG455 blocking filter for the
fourth transit, and the 300 lines/mm grism blazed at $26.7^\circ$ (300-26.7) with the
GG455 blocking filter for the fifth transit.

On average, these setups provide a
resolving power of $R\sim1200$, or approximately \SI{4.7}{\angstrom} per resolution
element, and access a full wavelength coverage of \SIrange{5050}{9840}{\angstrom}. In
practice, the signal"-to"-noise ratio (SNR) blueward of \SI{\approx 5400}{\angstrom} and
redward of \SI{\approx 9300}{\angstrom} dropped to roughly less than 25\% of peak counts.
For this reason, we omitted measurements outside of this range for the rest of the study.
Finally, we omitted data taken during twilight and at airmasses of $Z \ge 2$, where large
deviations in the white"-light curve became apparent.

We summarize our observing configurations and conditions for each night in
Table~\ref{tab:obs_log}. We also note that the difference in wavelength coverage between
nights comes from our choice to use wavelength ranges that maximized the integrated
flux for a given night over uniformity in wavelength coverage. This might produce slight
variations in the observed wavelength"-binned transit depths, which can contribute to
spurious slopes in the final transmission spectrum \citep{mcgruder2020}. For HAT-P-23b,
however, we recover a transmission spectrum with no typical activity-related slope (see
Section~\ref{sec:atmospheric_modeling}), so this does not appear to be a significant
effect in this case.

\begin{deluxetable*}{cccccccc}[htbp]
    \tablecaption{%
        Target and comparison star magnitudes and coordinates from
        \url{https://vizier.u-strasbg.fr/viz-bin/VizieR?-source=I/322}.
        \label{tab:comp_stars}
    }
    \tablehead{
        \colhead{Star} & \colhead{RA-J2000 (h:m:s)} & \colhead{Dec-J2000 (d:m:s)} &
        \colhead{B} & \colhead{V} & \colhead{J} & \colhead{K} &
        \colhead{D\tablenotemark{a}}
    }
    \startdata
    \hp{} & 20:24:29.7 & 16:45:43.8 & 13.4 & 12.1 & 11.1 & 10.8 & 0.00 \\
    comp4    & 20:23:29.4 & 16:49:08.0 & 13.5 & 12.5 & 10.7 & 10.0 & 0.50 \\
    comp5    & 20:24:15.1 & 16:42:08.4 & 13.8 & 12.7 & 11.0 & 10.4 & 0.36 \\
    \enddata
    \centering
    \tablenotetext{a}{%
        $D \equiv \sqrt{[(B-V)_c-(B-V)_t]^2 + [(J-K)_c - (J-K)_t]^2}$, where
        $c$ and $t$ refer to the given comparison star and HAT-P-23,
        respectively, in magnitudes.
    }
\end{deluxetable*}

\begin{deluxetable*}{CcccCcCCC}[htbp]
    \tablecaption{%
        Observation log\tablenotemark{a}
        \label{tab:obs_log}
    }
    
    \tablehead{
        \colhead{Transit} &
        \colhead{UT Date} &
        \colhead{Grism} &
        \colhead{BF} &
        \colhead{Mask (Cal, Sci)} &
        \colhead{Exp} &
        \colhead{SF} &
        \colhead{Airmass\tablenotemark{b}}
    }
    \startdata
    1 & 2016-06-22 06:13$\to$10:03 & 150-18.8 & WB5600 & 20  \times0.5,\ 20  \times12  & 15      & 300 & 1.5\to1.4\to2.2 \\
    2 & 2017-06-10 05:26$\to$09:08 & 300-17.5 & GG495  & 20  \times0.5,\ 90  \times10  & 50      & 165 & 1.9\to1.4\to1.5 \\
    3 & 2018-06-04 05:32$\to$10:26 & 300-17.5 & GG495  & 20  \times0.5,\ 90  \times10  & 30, 70  & 181 & 2.2\to1.4\to1.8 \\
    4 & 2018-06-21 05:14$\to$10:03 & 300-17.5 & GG455  & 20  \times0.5,\ 90  \times10  & 50, 70  & 172 & 1.7\to1.4\to2.1 \\
    5 & 2018-08-22 00:14$\to$06:47 & 300-26.7 & GG455  & 20  \times0.5,\ 90  \times10  & 70, 100 & 182 & 2.1\to1.4\to2.4
    \enddata
    \begin{flushleft}
    \tablenotetext{a}{%
        \textbf{Abbreviations:} Blocking Filter (BF), Calibration (Cal) and science (Sci)
            mask slit dimensions in arcseconds, Exposure Time in seconds (Exp), Number of
            science frames (SF).
    }
    \tablenotetext{b}{%
        The arrows indicate measurements taken at the beginning, middle, and end of the
            night.
    }
    \end{flushleft}
\end{deluxetable*}

\subsection{Data Collection} \label{ssec:data_collection}
We observed the first transit (Transit 1) on 22 Jun 2016, the second (Transit 2) on 10 Jun
2017, the third (Transit 3) on 04 Jun 2018, the fourth (Transit 4) on 21 Jun 2018, and the
fifth (Transit 5) on 22 Aug 2018. We used blocking filters for all nights to reduce
contamination from light at higher orders, while also truncating the spectral range to
\SIrange{5300}{9200}{\angstrom}. We made observations in Multi-Object Spectroscopy mode
with $2{\times}2$ binning in Fast readout mode (31\,s) for all nights to take advantage of
the reduced readout noise. Depending on the instrument setup and seeing conditions, we
adjusted the individual exposure times between \SIrange{15}{100}{} seconds to keep the
number of counts per pixel between \SIrange{30000}{49000}{} counts (ADU; gain = 1e$^-$/ADU
on f/2 camera), i.e., within the linearity limit of the CCD \citep{bixel2019}.

With the calibration mask in place, we took a series of arcs using a HeNeAr lamp before
each transit time"-series observation. The narrower slit width of the calibration mask
increased the spectral resolution of the wavelength calibration as well as avoided
saturation of the CCD from the arc lamps. We took a sequence of high"-SNR flats with a
quartz lamp through the science mask to characterize the pixel"-to"-pixel variations in
the CCD. As in our past studies \citep{rackham2017, espinoza2019, bixel2019, weaver2020,
mcgruder2020}, we did not apply a flat"-field correction to the science images because it
introduces additional noise in the data and does not improve the final results.

During the Transit 5 observation, the incorrect grism (300-26.7) was mistakenly inserted
into the instrument in place of the intended 300-17.5 grism used in previous nights. This
led to the grism metadata stored in the header files to be incorrectly labelled as the
300-17.5 grism instead of 300-26.7, which we did not notice until after the data had been
collected and the spectra examined and compared with previous data. The end result is data
that are still usable, but slightly less dispersed relative to the other nights
($\SI{1.25}{\angstrom/\text{pixel}}$ vs. $\SI{1.341}{\angstrom/\text{pixel}}$), which we
account for in the rest of our analysis. We show representative science frames from the
CCD array in Figure~\ref{fig:raw_frames} of the Appendix.

\section{Data reduction and light curve analysis} \label{sec:reduction}
\subsection{Reduction pipeline} \label{ssec:reduction_pipeline}
We reduced the raw data using the ACCESS pipeline described in previous ACCESS
publications \citep{rackham2017, espinoza2019, bixel2019, weaver2020, mcgruder2020}. The
detailed functions of the pipeline, including standard bias and flat calibration, bad
pixel and cosmic ray correction, sky subtraction, spectrum extraction, and wavelength
calibration are described in \citet{jordan2013} and \citet{rackham2017}. We briefly
summarize the data reduction here.

We applied the wavelength solution found with the arc lamps to the first science image
spectra and the remaining image spectra were then cross"-correlated with the first image.
We calibrated the spectra from all stars to the same reference frame by identifying shifts
between the \halpha{} absorption line minimum of the median spectra and air wavelength of
\halpha{}, and interpolated the spectra onto a common wavelength grid using b"-splines. We
aligned all spectra to within \SI{2}{\angstrom}, which is less than our average resolution
element of \SI{4.7}{\angstrom} (see Section~\ref{ssec:general_setup}).

\begin{figure*}[htbp]
\gridline{
    \fig{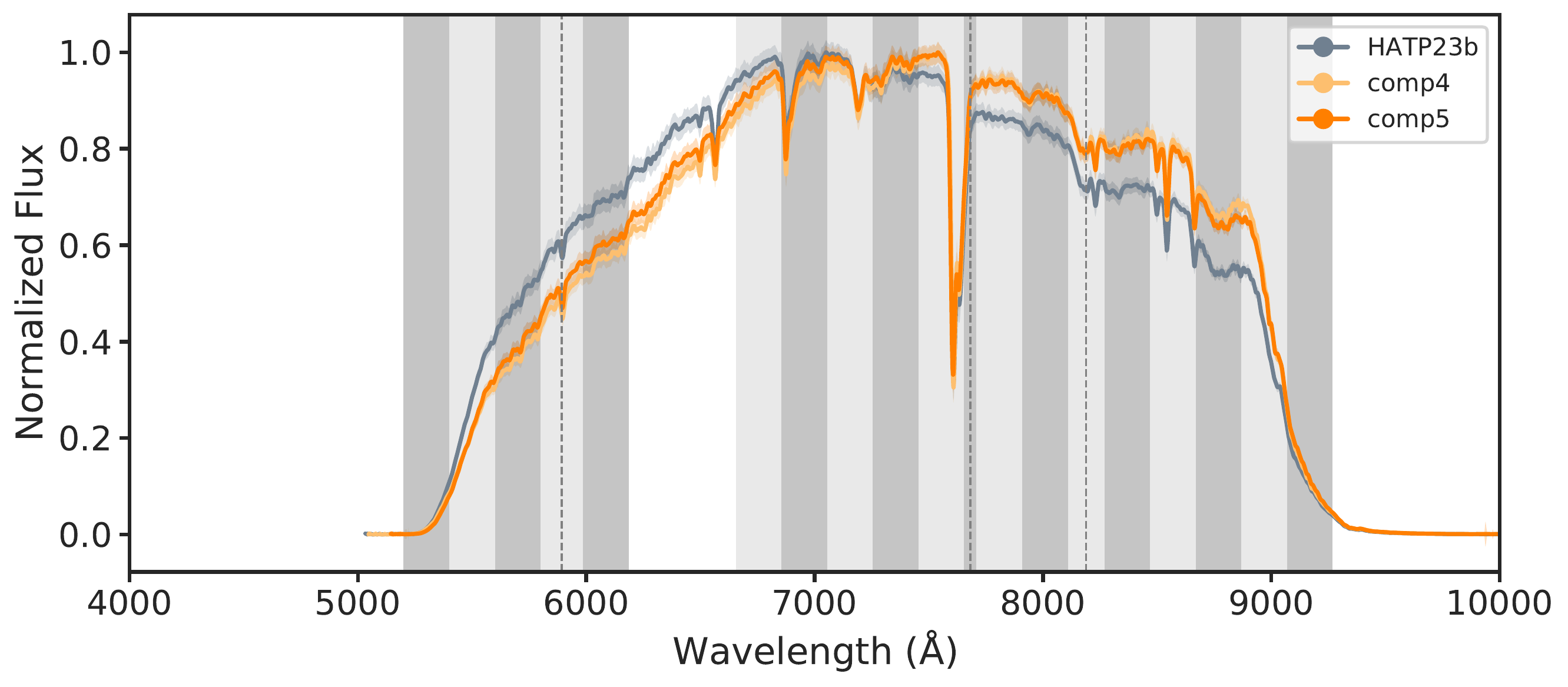}{0.49\linewidth}
        {(a) Transit 1}
}
\gridline{
    \fig{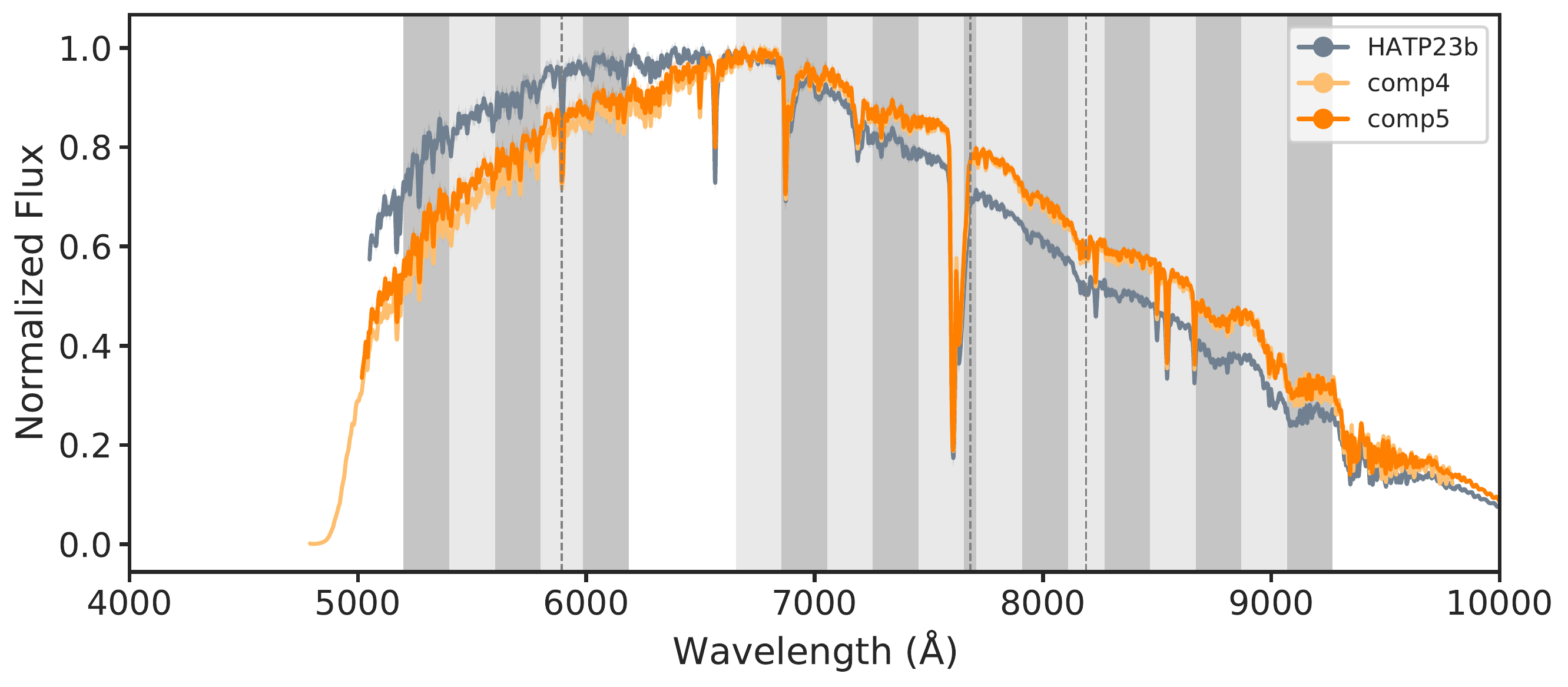}{0.49\linewidth}
        {(b) Transit 2}
    \fig{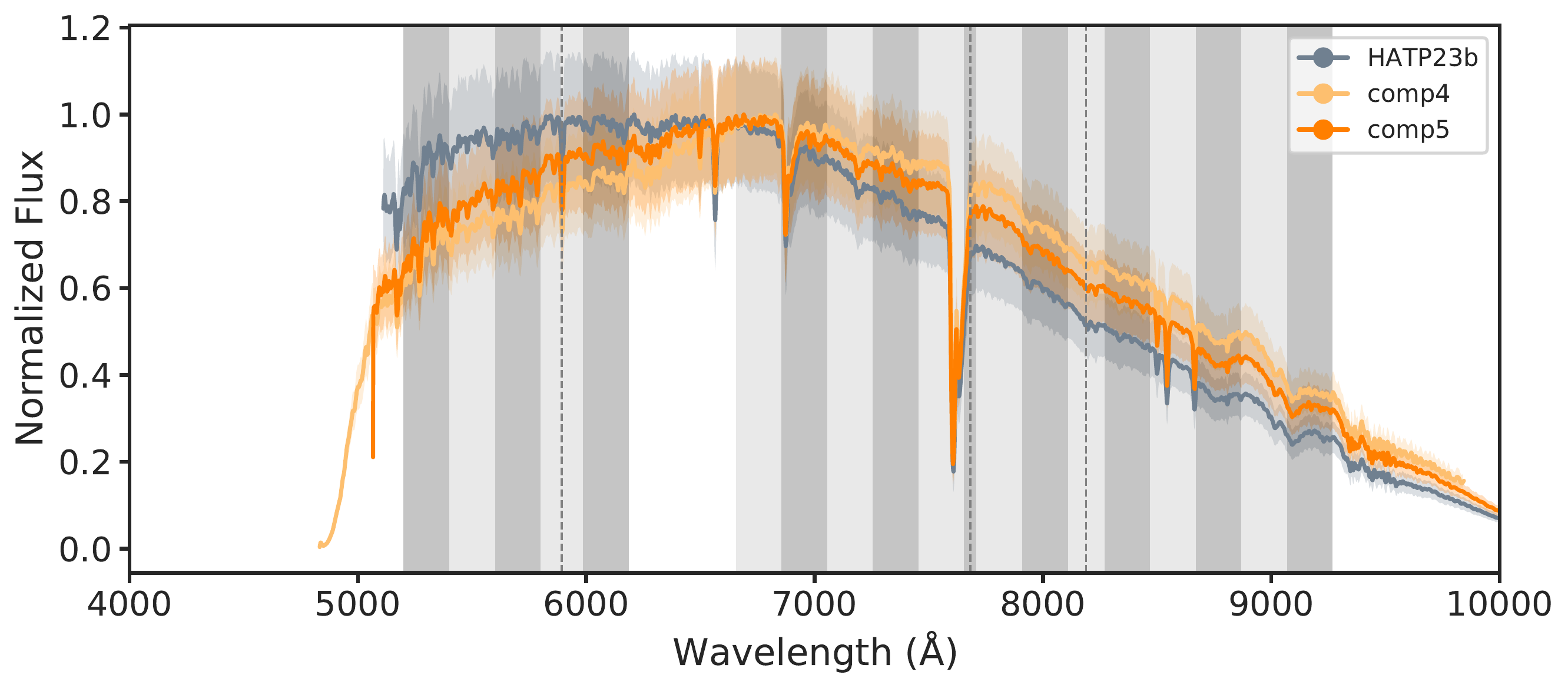}{0.49\linewidth}
        {(c) Transit 3}
}
\gridline{
    \fig{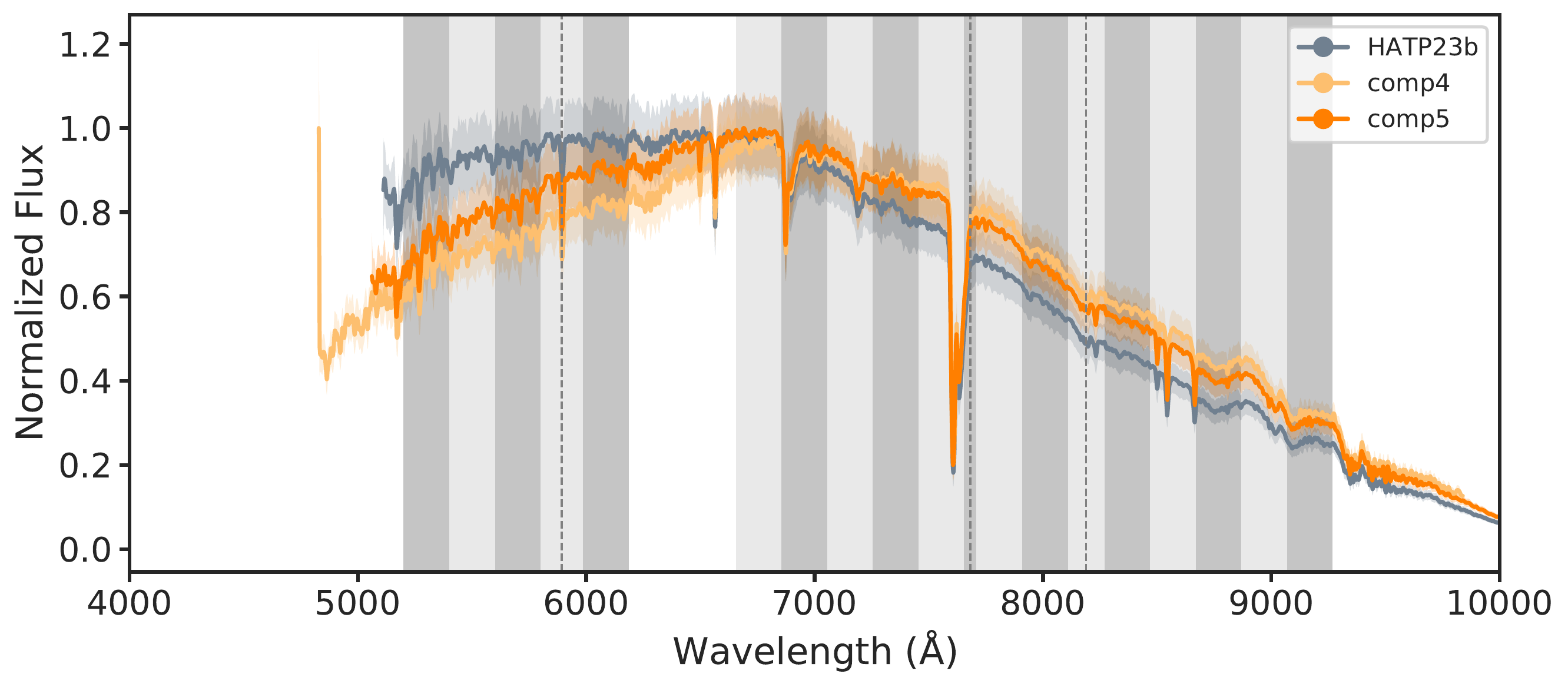}{0.49\linewidth}
        {(d) Transit 4}
    \fig{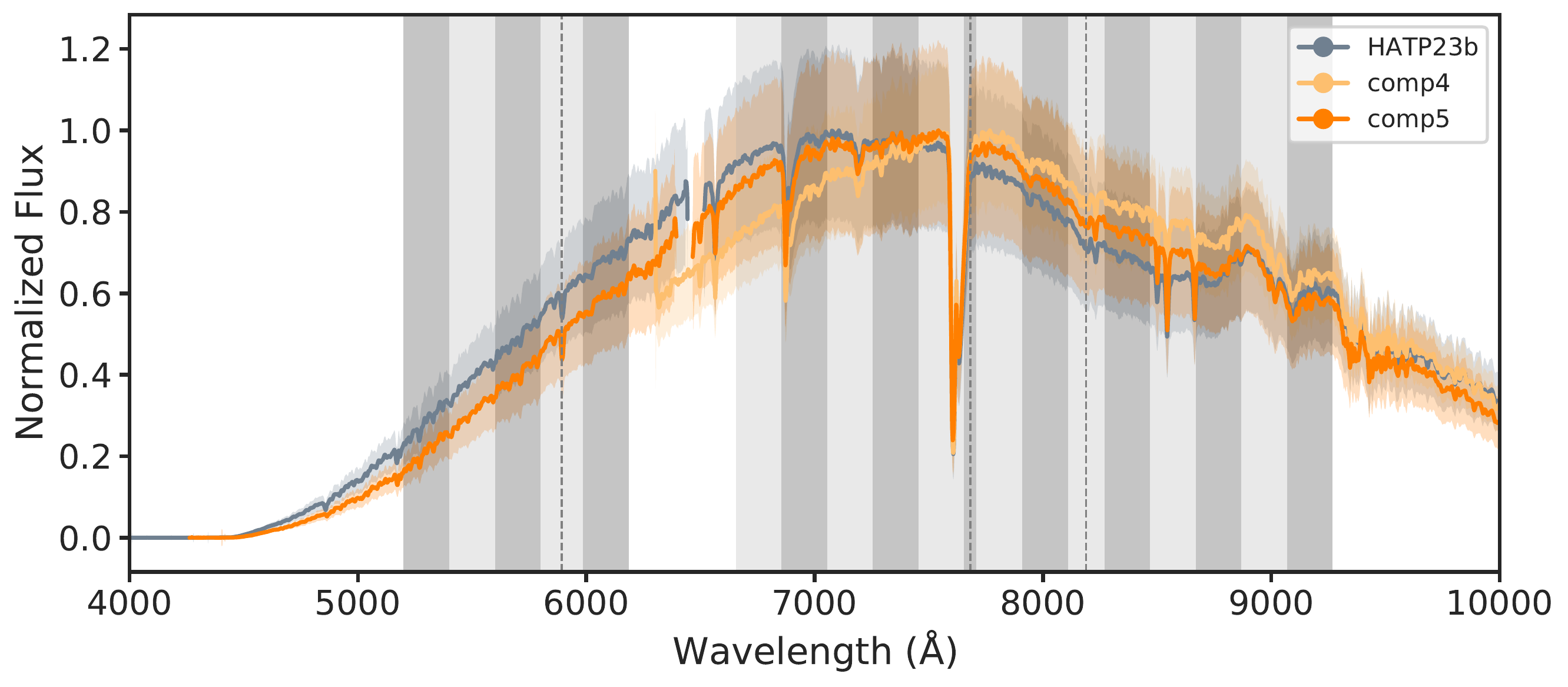}{0.49\linewidth}
        {(e) Transit 5}
}
\caption{%
        Optimally extracted spectra over the course of each night for each star. The
        median (solid line) and $1\sigma$ region region are shown for the target and each
        comparison star used, along with the wavelength bin scheme used. We note that the
        first and last bin needed to be omitted for Transit 1 due to lack of flux. The gap
        in bins is due to the chip gap introduced in Transit 5.
        \pylink{https://icweaver.github.io/HAT-P-23b/notebooks/extracted_spectra.html}
}
\label{fig:extracted_spectra}
\end{figure*}

The final results of the pipeline are sets of wavelength"-calibrated and extracted spectra
for the target and each comparison star for each night as shown in
Figure~\ref{fig:extracted_spectra}. We use those spectra to produce integrated
white"-light (Figure~\ref{fig:extracted_wlcs} in the Appendix) and spectrally binned light
curves. The series of white"-light curves produced in this fashion informed which
comparison stars to omit in the rest of the analysis on a per transit basis, which we
describe in Section~\ref{ssec:comp_star_selection}.

\subsection{Wavelength binning} \label{ssec:bin_schemes}
For our main analysis, we used a set of semi"-uniform bins with a main width of
\SI{200}{\angstrom}, which we found to be a good compromise between spectral resolution
and signal"-to"-noise considerations. In addition, we interspersed narrower bins centered
around the air wavelength locations of \Na{}"-D, \K, and \Na{}"-8200 with widths of
\SI{185.80}{\angstrom}, \SI{54.00}{\angstrom}, and \SI{160.80}{\angstrom}, respectively.
We made the \K bin relatively narrow to avoid nearby tellurics. To check the results from
our analysis based on this wavelength binning scheme, we separately used a set of 5 bins
centered around each of these three species, set by the full width of their respective
absorption lines (\SI{45}{\angstrom}, \SI{10}{\angstrom}, and \SI{40}{\angstrom}). We
share our two wavelength bin schemes in \mbox{Tables \ref{tab:bins} and \ref{tab:species}}
of the Appendix.

\subsection{Comparison star selection} \label{ssec:comp_star_selection}
For each transit, we divided the white"-light curve of \hp{} by each of the comparison
stars common to each night (comp4 and comp5) to select which data points to use for each
night (Figure~\ref{fig:extracted_wlcs}). For a given comparison star, we defined fluxes
more than $2\sigma$ apart from adjacent points in the divided white"-light curve to be
potential outliers for that night, where $1\sigma$ was defined to be the photometric noise
for that night. Because this set could contain false"-positives caused by an actual
non"-outlying point surrounded by two positive outliers, we then manually went back and
removed these types of points from the potential set of outliers. Conversely, we also
manually included false-negatives in the outlier set. These were points not initially
marked as outliers because they were part of an overall feature, for example the large
decrease in flux. Finally, we omitted the union of these selected outliers from all nights
from the rest our of analysis. With this set of selected data points and comparison stars
for each transit epoch, we next performed the detrending of the resulting light curves as
described in Section~\ref{sec:detrending}.

\section{Detrending analysis}\label{sec:detrending}
\subsection{Methodology overview}
We used Gaussian Process (GP) regression techniques combined with principle component
analysis (PCA) to detrend our reduced data, following \citep{espinoza2019}. GPs are a
flexible and robust tool for modeling data in the machine learning community
\citep{rasmussen2005} that has gained popularity in the exoplanets field (see, e.g.,
\citealt{gibson2012}, \citealt{aigrain2012}). \citet{gibson2012} provides a good overview
to this methodology applied to exoplanet transit light curves. Applying this methodology
for a collection of $N$ measurements $(\bm f)$, such as the flux of a star measured over a
time series, the log marginal likelihood of the data can be written as:
\begin{align}
    \log\mathcal L(\bm r | \textbf{\textsf X},\bm\theta,\bm\phi)
    = &-\frac{1}{2}\bm r^\top\Sigma^{-1}\bm r
    -\frac{1}{2}\log\left|\Sigma\right| \\
      &-\frac{N}{2}\log(2\pi)\quad,
\end{align}
where $\bm r \equiv \bm f - T(\bm t, \bm\phi)$ is the vector of residuals between the data
and analytic transit function $T$; $\textbf{\textsf X}$ is the $N\times K$ matrix for $K$
additional parameters, where each row is the vector of measurements
$\bm x_n = (x_{n,1}, \cdots x_{n,K})$ at a given time $n$; $\bm\theta$ are the
hyperparameters of the GP; $\bm\phi$ are the transit model parameters; and $\Sigma$ is the
covariance of the joint probability distribution of the set of observations $\bm f$. In
our analysis, we used six systematics parameters: time, full width at half maximum (FWHM)
of the spectra on the CCD, airmass, position of the pixel trace through each spectrum on
the chip of the CCD, sky flux, and shift in wavelength space of the trace.

We used the Python package \texttt{batman} \citep{batman} to generate our analytic
transit model. We used the open-source package
\texttt{ld-exosim}\footnote{\url{https://github.com/nespinoza/ld-exosim}}
\citep{espinoza_jordan_16} to identify the appropriate limb-darkening law for this model,
which we found to be the logarithmic law. From here, the log posterior distribution
$\log\mathcal P(\bm\theta,\bm\phi|\bm f,\textbf{\textsf X})$ can be determined by placing
explicit priors on the maximum covariance hyperparameters and the scale length
hyperparameters. From $\mathcal P$, the transit parameters can then be inferred by
optimizing with respect to $\bm\theta$ and $\bm\phi$.

\subsection{Sampling methodology}
We accomplished the above optimization problem with the Bayesian inference tool,
\texttt{PyMultiNest} \citep{buchner2014}, using 1000 live points and computed the log
likelihoods from the GP with the \texttt{george} \citep{george} package. We implemented
this detrending scheme by simultaneously fitting the data with a Matérn"-3/2 kernel for
the GP under the assumption that points closer to each other are more correlated than
points farther apart. The PCA methodology follows from \citet{jordan2013} and
\citet{espinoza2019}, where $M$ signals, $S_i(t)$, can be extracted from $M$ comparison
stars and linearly reconstructed according to the eigenvalues, $\lambda_i$, of each
signal. This allows for the optimal extraction of information from each comparison star to
inform how the total flux of \hp{} varies over the course of the night.

\subsection{Parameterization improvements}
In past work, we depended on the ratio of the semi"-major axis to stellar radius $(a/R_s)$
as an input transit parameter. With \textit{Gaia}, we are now able to invert the problem
by using the newly determined measurements of the stellar density $(\rho_s)$ from
\textit{Gaia} Data Release 2 \citep{GAIA2018} to place tighter constraints on $a/R_s$.
Applying our standard assumption of a circular Keplerian orbit with period $P$ then yields
$a/R_s = [(GP^2)/(3\pi)]^{1/3}\rho_s^{1/3}$, where $G$ is the gravitational constant. In
addition, we implemented the technique introduced by \citet{espinoza2018} for
reparameterizing the impact parameter $(b)$ and planet"-to"-stellar radius ratio $(p
\equiv \RpRs)$ to further improve our transit parameter sampling. In this technique, these
transit parameters are expressed as functions of the random variables $r_1 \sim U(0, 1)$
and $r_2 \sim U(0, 1)$, and efficiently sampled from a region bounded by the physical
constraints of transiting systems (i.e. $0 < b < 1 + p$). This is a natural analog to the
uniform sampling method of limb darkening coefficients introduced in \citet{kipping2013}.

\subsection{Transmission spectrum construction}
\begin{figure*}[htbp]
\gridline{
    \fig{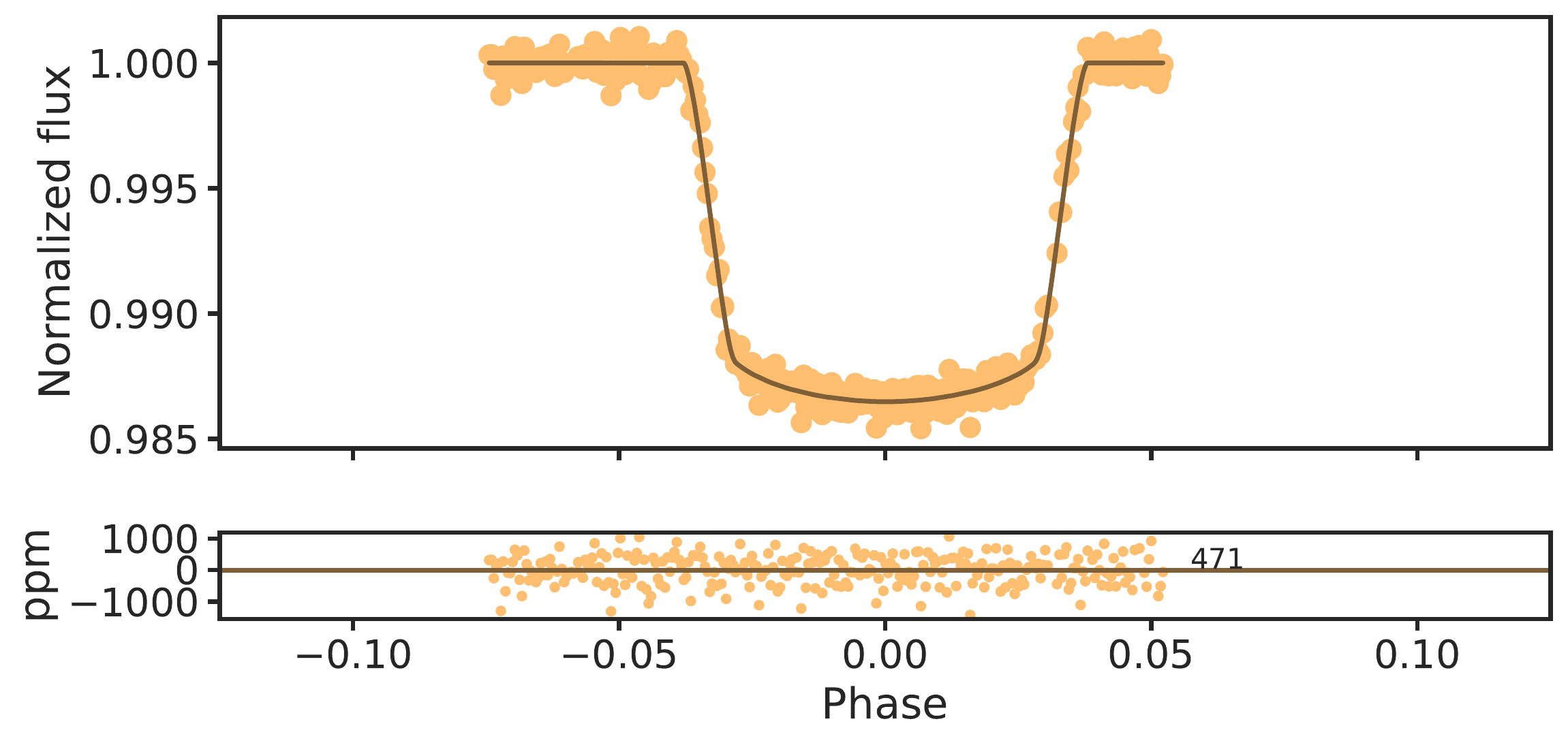}{0.49\linewidth}
        {(a) Transit 1}
}
\gridline{
    \fig{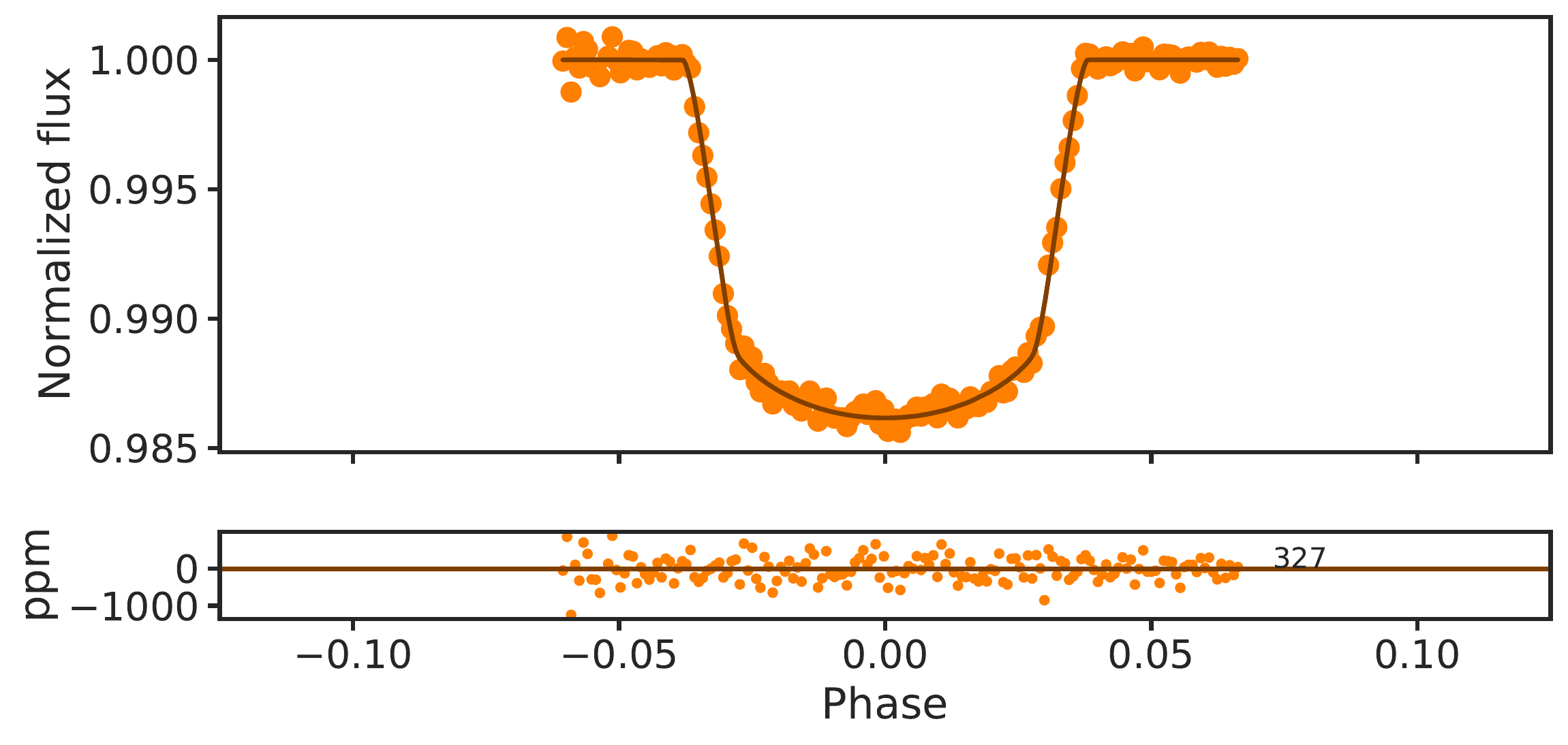}{0.49\linewidth}
        {(b) Transit 2}
    \fig{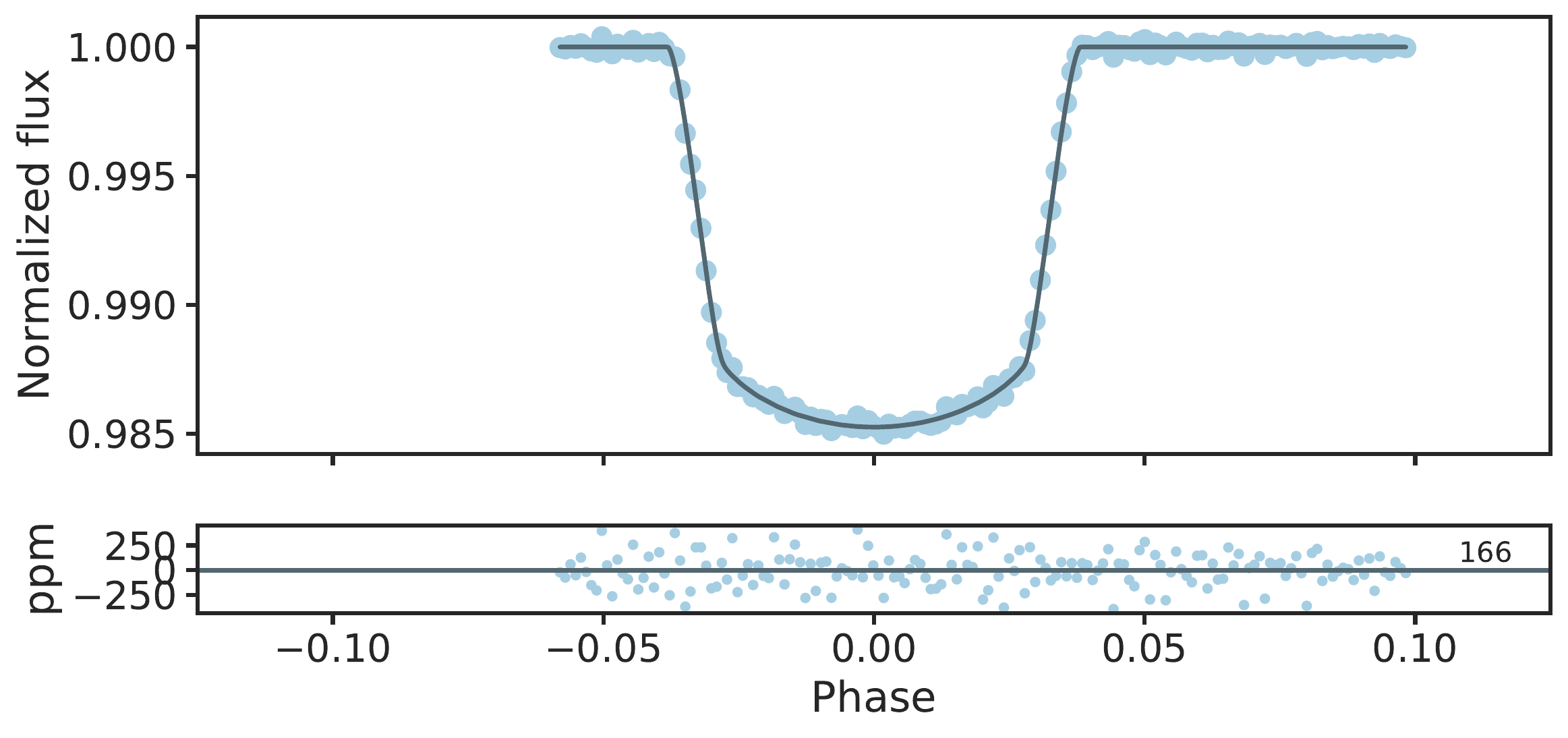}{0.49\linewidth}
        {(c) Transit 3}
}
\gridline{
    \fig{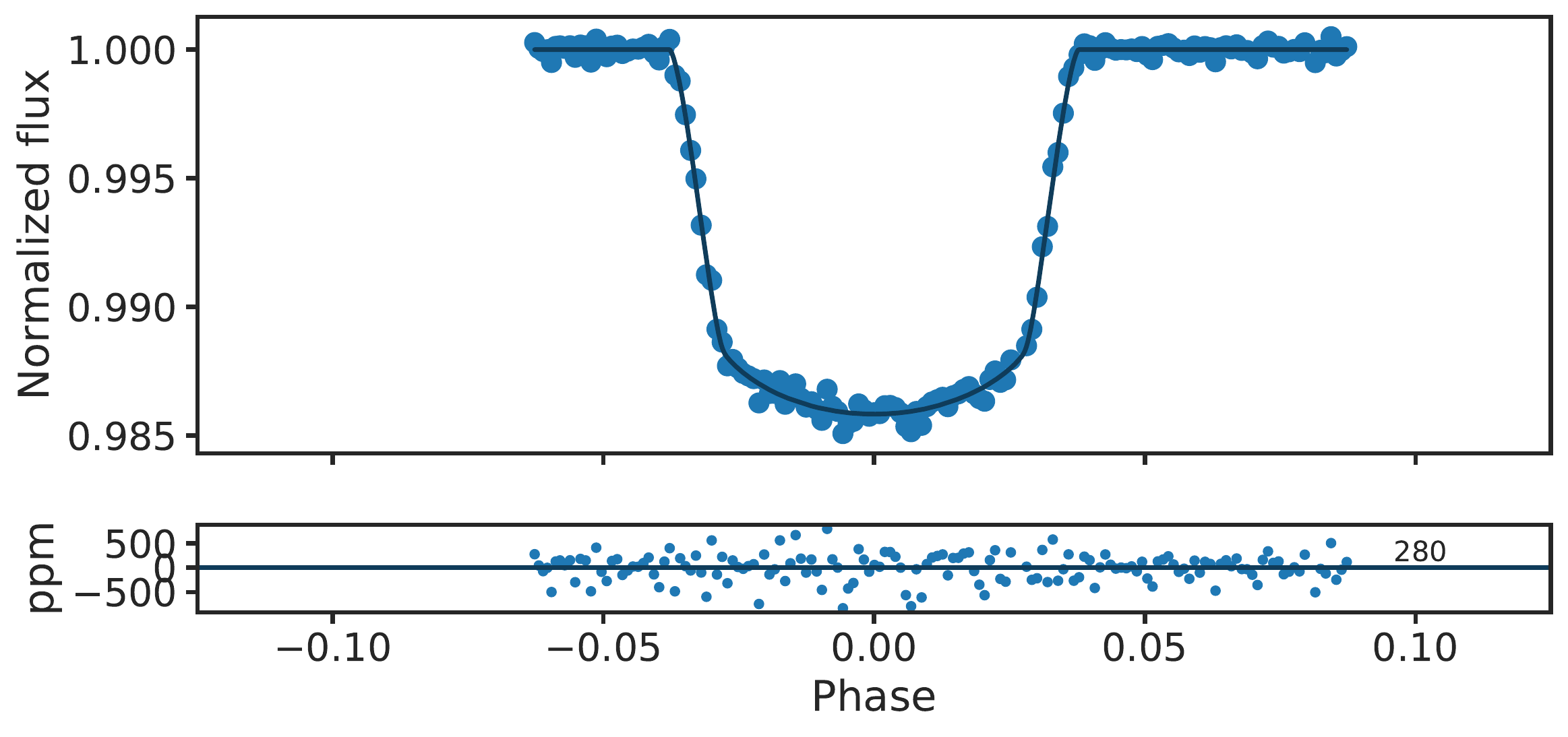}{0.49\linewidth}
        {(d) Transit 4}
    \fig{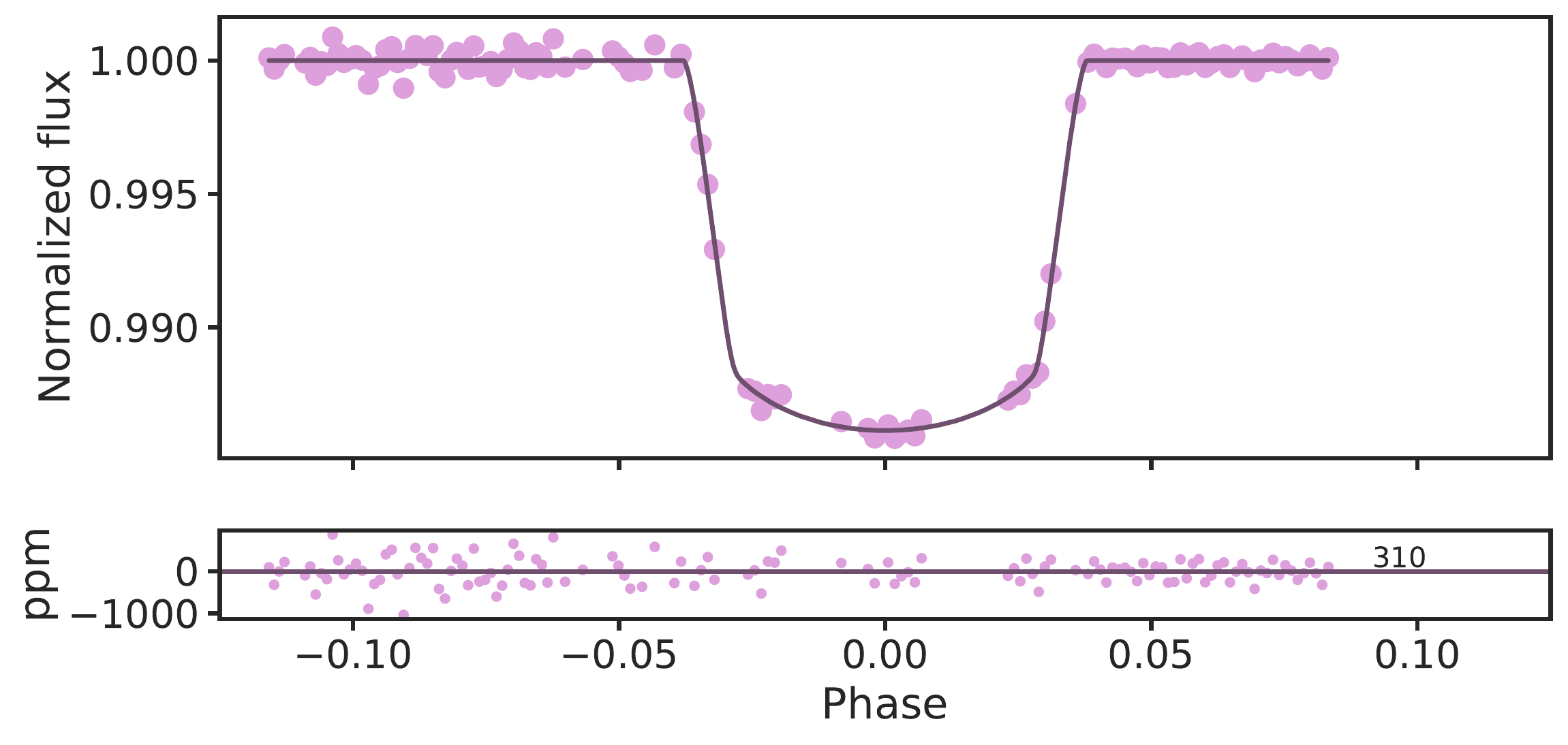}{0.49\linewidth}
        {(e) Transit 5}
}
\caption{%
        Detrended white"-light curves for each night used in our final analysis. The
        white"-light curves are shown for the maximum number of principal components used
        (2). We share the model"-averaged transit parameters in
        Table~\ref{tab:detrended_wlcs} and associated corner plots in
        Figure~\ref{fig:detrended_wlcs_corners} of the Appendix.
        \pylink{https://icweaver.github.io/HAT-P-23b/notebooks/detrended_wlcs.html}
}
\label{fig:detrended_wlcs}
\end{figure*}

Finally, we Bayesian model averaged the principal components together, which we determined
by detrending with one, then two, up to $M$ principal components, to create the final
detrended white"-light curves and model parameters of interest. We present the resulting
white"-light curves in Figure~\ref{fig:detrended_wlcs}, the best"-fit parameters in
Table~\ref{tab:detrended_wlcs}, and associated corner plots in
Figure~\ref{fig:detrended_wlcs_corners} of the Appendix. To create the transmission
spectrum for each night (Figure~\ref{fig:tspec_full}), we applied the same detrending
methodology to the wavelength-binned light curves (described in
Section~\ref{ssec:bin_schemes}), keeping all transit parameters fixed to the fitted
white"-light curve values except for the transit depth and limb darkening parameters.
Because the impact parameter needed to be held fixed for the binned fitting, we sampled directly for it instead of uniformly sampling in $(b, p)$ space. We used a truncated normal distribution for this sampling, given by:
\begin{align}
\xi &= \frac{x - \mu}{\sigma},\enskip \alpha = \frac{a-\mu}{\sigma},\enskip \beta = \frac{b-\mu}{\sigma} \\
Z &= \Phi(\beta) - \Phi(\alpha) \quad,
\end{align}
where $x$ is the binned planet-to-star radius ratio, $\mu$ is the mean fitted white"-light curve depth, $\sigma$ is the corresponding $1\sigma$ uncertainty on that depth, $a=0$, $b=1$, and $\Phi(\cdot)$ is the cummulative distribution function.

\begin{deluxetable*}{cRRRRR}[htbp]
%\tabletypesize{\scriptsize}
\tablecaption{%
    Fitted model averaged transit parameters from GP+PCA detrending method shown in
    Figure~\ref{fig:detrended_wlcs}. We share the associated corner plots in
    Figure~\ref{fig:detrended_wlcs_corners} of the Appendix.
    \label{tab:detrended_wlcs}
}
\tablehead{
    \colhead{parameter\tablenotemark{a}} &
    \colhead{Transit 1} &
    \colhead{Transit 2} &
    \colhead{Transit 3} &
    \colhead{Transit 4} &
    \colhead{Transit 5}
}
\startdata
$\delta$                    & 12584.79890^{+633.98979}_{-647.93827} & 12716.26467^{+421.08962}_{-459.31331} & 13571.18577^{+598.68259}_{-650.42694} & 13204.48610^{+472.94430}_{-465.37117} & 12831.57340^{+456.65963}_{-489.88214} \\
$R_\mathrm{p}/R_\mathrm{s}$ & 0.11218^{+0.00283}_{-0.00289}         & 0.11277^{+0.00187}_{-0.00204}         & 0.11650^{+0.00257}_{-0.00279}         & 0.11491^{+0.00206}_{-0.00202}         & 0.11328^{+0.00202}_{-0.00216}         \\
$t_0$                       & 4852.26537^{+0.00015}_{-0.00015}      & 4852.26540^{+0.00014}_{-0.00014}      & 4852.26541^{+0.00014}_{-0.00013}      & 4852.26545^{+0.00015}_{-0.00015}      & 4852.26543^{+0.00013}_{-0.00014}      \\
$P$                         & 1.21289^{+1.13e-07}_{-1.14e-07}       & 1.21289^{+8.45e-08}_{-8.76e-08}       & 1.21289^{+8.76e-08}_{-8.78e-08}       & 1.21289^{+7.42e-08}_{-7.84e-08}       & 1.21289^{+7.00e-08}_{-7.13e-08}          \\
$\rho_\mathrm{s}$           & 1.07230^{+0.08406}_{-0.07603}         & 0.96894^{+0.06463}_{-0.05683}         & 1.01504^{+0.05602}_{-0.05063}         & 1.03666^{+0.06581}_{-0.06079}         & 1.02626^{+0.07229}_{-0.07204}         \\
$i$                         & 84.33384^{+0.88921}_{-0.80101}        & 83.14471^{+0.73406}_{-0.61691}        & 83.64261^{+0.56957}_{-0.51338}        & 83.76319^{+0.69346}_{-0.64068}        & 83.73923^{+0.73109}_{-0.74358}        \\
$b$                         & 0.43173^{+0.04848}_{-0.05892}         & 0.50417^{+0.03445}_{-0.04412}         & 0.47517^{+0.02969}_{-0.03537}         & 0.46904^{+0.03843}_{-0.04320}         & 0.46936^{+0.04278}_{-0.04503}         \\
$a/R_\mathrm{s}$            & 4.36890^{+0.11130}_{-0.10580}         & 4.22377^{+0.09190}_{-0.08424}         & 4.28971^{+0.07750}_{-0.07254}         & 4.31996^{+0.08955}_{-0.08614}         & 4.30547^{+0.09880}_{-0.10320}         \\
$u$                         & 0.24859^{+0.10290}_{-0.10754}         & 0.38699^{+0.07698}_{-0.07472}         & 0.37032^{+0.10186}_{-0.11732}         & 0.40126^{+0.08643}_{-0.09298}         & 0.32686^{+0.07151}_{-0.08796}
\enddata
\centering
\tablenotetext{a}{%
    \textbf{Parameter definitions:}
    $\delta = (\RpRs)^2$ - transit depth (ppm),
    $\RpRs$ - planet radius / stellar radius, \\
    $t_0$ mid-transit time (JD - 245000),
    $P$ - period (days),
    $\rho_s$ - stellar density (g/cc),
    $i$ - inclination (degrees), \\
    $b$ - impact parameter,
    $a/R_s$ - semi-major axis / stellar radius,
    $u$ - linear limb darkening coefficient.
}
\end{deluxetable*}

\section{Stellar Activity} \label{sec:stell_act}
Before combining the transmission spectra from each night, we first considered the impact
of stellar photospheric heterogeneity, which can have an observable effect on transmission
spectra \citep{pont2008, pont2013, sing2011, oshagh2014, zhang2018}, even if magnetically
active regions are not occulted by the transiting exoplanet \citep{mccullough2014,
rackham2018, rackham2019, apai2018}. Qualitatively, global variations in stellar activity
could manifest themselves as an overall dimming or brightening of the star, which could
lead to significant variations in transit depths. Changes in photometric activity can
roughly correlate with the covering fraction of starspots, which in turn can modulate the
luminosity of the star and impact observed transit depths \citep{berta2011}, although
\citet{rackham2019} show that this relationship is very non-linear. Those variations can
also be wavelength"-dependent, leading to slopes with spurious spectral features in the
transmission spectrum.

To assess the brightness variations of \hp{} over the time frame of our observations, we
used \mbox{2015 Aug 12 -- 2018 Oct 17} photometric observations from 566 out"-of"-transit
V"-band images of \hp{} taken by Ohio State University's All"-Sky Automated Survey for
Supernovae\footnote{\url{https://asas-sn.osu.edu}} (ASAS"-SN) program
\citep{shappee2014,kochanek2017}.

\begin{figure*}[htbp]
    \centering
    \includegraphics[width=0.98\linewidth]{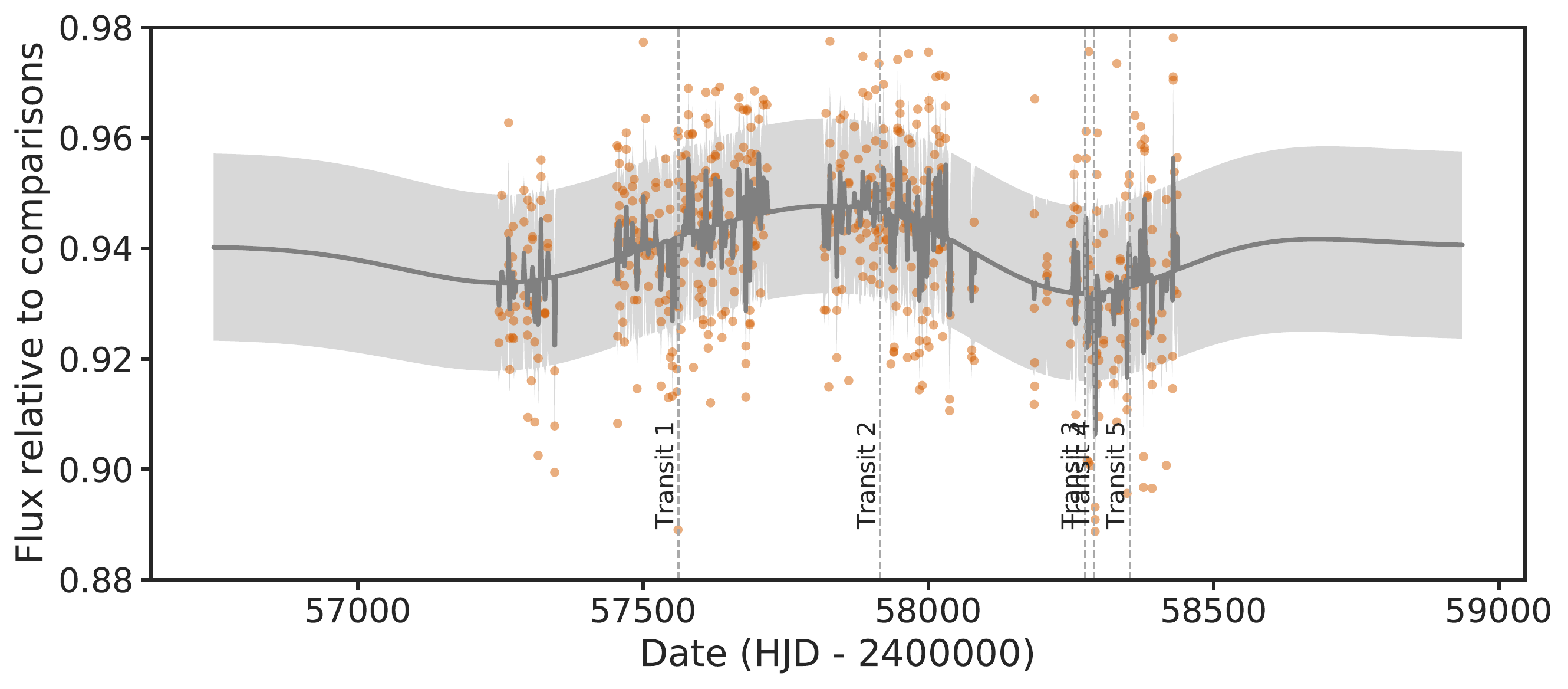}
    \caption{%
        Ground"-based photometric observations of \hp{} from ASAS"-SN (points) during each
        transit epoch (dashed vertical lines). The flux is relative to the average
        brightness of the comparison stars used in this survey. The Gaussian process
        regression model (solid line) and $1\sigma$ uncertainty (shaded region) fit to
        ASAS"-SN data are also overplotted.
        \pylink{https://icweaver.github.io/HAT-P-23b/notebooks/photometric_act.html}
    }
\label{fig:photometric_act}
\end{figure*}

The ASAS"-SN photometry is sampled much more coarsely than our transit observations, so we
again used a GP routine \citep{carter2020, weaver2020, alam2018} to model quasi-periodic modulations in the data and
estimate the amplitude of the photometric variation induced by stellar activity during
each of the five transit epochs of \hp. Following \citet{alam2018}, we used a negative log
likelihood kernel along with a gradient"-based optimization routine to find the best"-fit
hyperparameters, assuming a stellar rotation period $P_\text{rot} = \SI{7}{days}$ from
\citet{schrijver2020}. Figure~\ref{fig:photometric_act} shows the GP regression model for
the relevant ASAS"-SN data. Overall, the relative flux from the photometric monitoring
model varied by as much as 2\% from the median value obtained from the GP. Although the
ASAS"-SN observations are too coarse to effectively sample the photometric activity during
times of transits, this data still gives us valuable insight into the variations in
stellar flux between epochs.

The white"-light transit depths (Table~\ref{tab:detrended_wlcs}) between these epochs
varied by as much as \SI{600}{ppm} ($\approx 5\%$) from the weighted-mean white"-light
curve depth (\MeanWLCDepthErr), an effect that can be attributed to the observed stellar
variability, differences in the relative spectral flux of the comparison stars used, or
differences in wavelength coverage introduced by chip gaps \citep{mcgruder2020}.

We find that, in this case, changes in stellar disk coverage by unocculted heterogenieties
likely drive the white"-light curve depth variations that we observe between transit
epochs of HAT-P-23b. For example, assuming that unocculted starspots effectively radiate
as a single black body with temperature $T_\text{sp}$, the required covering fraction
$f_\text{sp}$ needed to produce the observed change in luminosity $(\Delta L_\text{obs}
\equiv L_\text{obs,sp} / L_\text{obs,0})$ would be:
\begin{align}
        f_\text{sp} &= \frac{T_0^4}{T_0^4 - T_\text{sp}^4}
        \left(1 - \Delta L_\text{obs}\right)\quad,
        \label{eqn:covering_fraction}
\end{align}
where $T_0$ is the effective temperature of HAT-P-23, $L_\text{obs,sp}$ is its observed
luminosity in the presence of the spot, and $L_\text{obs,0}$ is the observed luminosity
assuming an immaculate photosphere. Because we take this as a ratio, it is equivalent to
using the observed flux. Using the minimum relative flux and median relative flux
estimated from our GP analysis in Figure~\ref{fig:photometric_act} for this ratio
($\approx$ 0.92 and 0.94, respectively), the effective temperature of HAT-P-23b
$(T_0 \approx \SI{5900}{K})$, and temperatures for the spot $T_\text{sp} \approx$
\SIrange{2200}{3800}{K} (we justify this choice of temperatures in
Section~\ref{sec:atmospheric_modeling}), we estimate spot covering fractions ranging from
$\approx$ \SIrange{2.2}{2.6}{\%}. For this reason, we calculate and apply transit depth
offset corrections as described in Section~\ref{sec:tspec} before building the final
transmission spectrum. In Section~\ref{sec:atmospheric_modeling} we model the contribution
of an unocculted heterogeneous photosphere to the resulting transmission spectrum,
informed by the photometric monitoring data.

\section{Transmission Spectrum} \label{sec:tspec}
\subsection{Combining nights} \label{ssec:comb_nights}
The uncertainties in the wavelength-binned transit depths from our individual transits
(Table~\ref{tab:tspec_full}) range from \SIrange{208}{1445}{ppm}, which is not precise
enough to detect the atmosphere of \hpb{}. For example, a hydrogen"-dominated composition
for the atmosphere of \hpb{} would produce a signal $\Delta D$ of \PlanetSignal at 5 scale
heights \citep[based on eq.~11 of][ using the most up to date system parameters from the
literature shared in Section~\ref{sec:intro}]{Miller_Ricci_2009}. Therefore, we needed to
combine the transmission spectra from the five transits to be sensitive to atmospheric
features of the planet.

\begin{figure*}[htbp]
    \centering
    \includegraphics[width=0.98\linewidth]{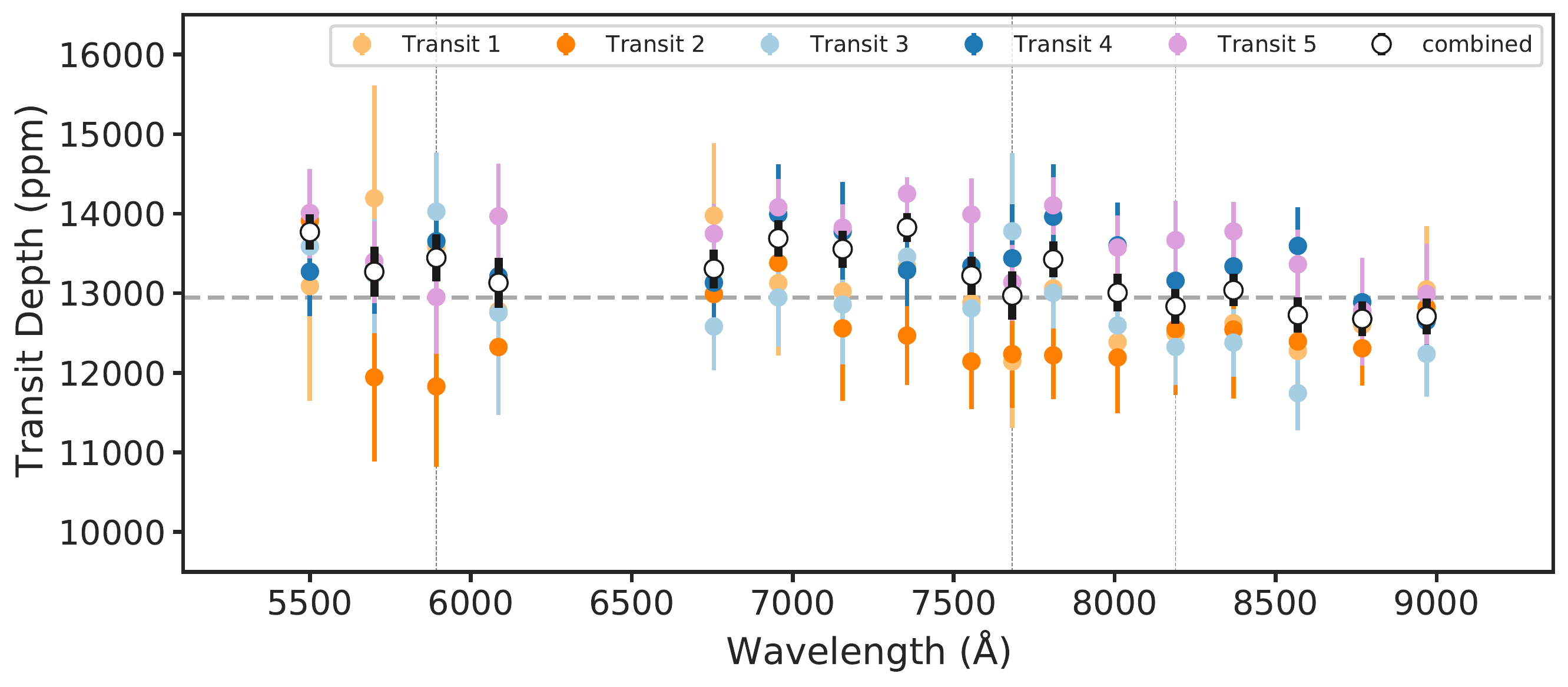}
    \caption{%
        Transmission spectrum built from binned light curves show in
        \mbox{Figures~\ref{fig:detrended_binned_lcs_transit_1} --
        \ref{fig:detrended_binned_lcs_transit_5}}. From left to right, the vertical dashed
        lines mark the air wavelength locations of \Na"-D, \K, and \Na"-8200,
        respectively. For reference, the weighted mean white"-light curve depth of
        \MeanWLCDepth{} is shown by the horizontal dashed line. The tabulated data for the
        above transmission spectra are available in Table~\ref{tab:tspec_full} of the
        Appendix.
        \pylink{https://icweaver.github.io/HAT-P-23b/notebooks/tspec_full.html}
    }
    \label{fig:tspec_full}
\end{figure*}

We offset each transmission spectrum by the difference in its white"-light curve depth
from the weighted mean white"-light curve depth for all nights and then average the offset
transmission spectra together, weighted by the wavelength"-dependent uncertainty estimated
from the wavelength"-binned fitting. We combined the asymmetrical uncertainties according
to Model 1 of \citet{barlow2003} and present the final results in
Figure~\ref{fig:tspec_full}.

\begin{figure*}[htbp]
    \centering
    \includegraphics[width=0.98\linewidth]{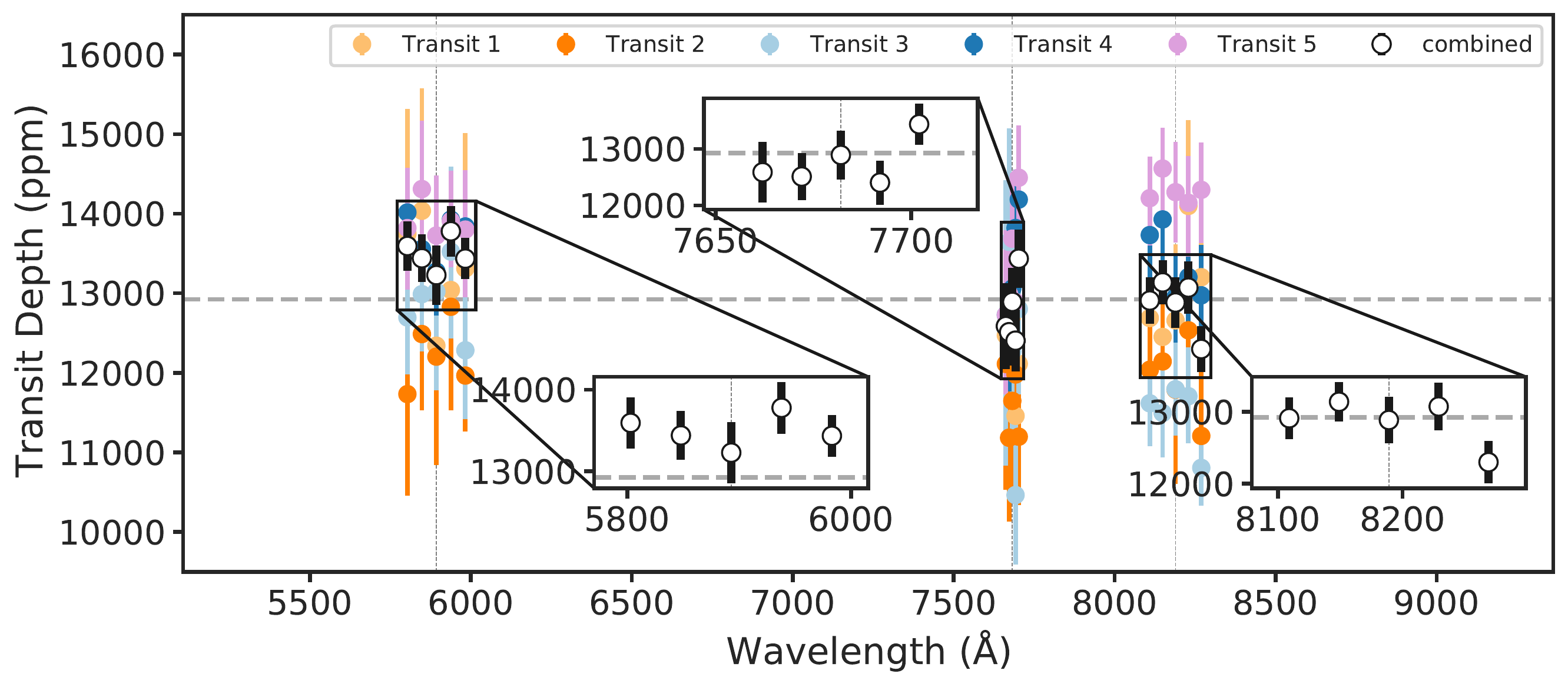}
    \caption{%
        Same as Figure~\ref{fig:tspec_full}, but for wavelength bins centered around the
        above species and with a resulting weighted mean transit depth of \MeanWLCDepthSp.
        We observe no excess absorption near any of the species investigated. The
        tabulated data for the above transmission spectra are available in
        Table~\ref{tab:tspec_species} of the Appendix.
        \pylink{https://icweaver.github.io/HAT-P-23b/notebooks/tspec_species.html}
    }
    \label{fig:tspec_species}
\end{figure*}

To verify our methodology, we also repeated our analysis up to this point using the
wavelength binning scheme described in Section~\ref{ssec:bin_schemes} that is centered
around \Na{}"-D, \K, and \Na{}"-8200. The resulting combined transmission spectrum
(Figure~\ref{fig:tspec_species}) shows no excess absorption near the species we examined,
in agreement with our retrieval analysis discussed in
Section~\ref{sec:atmospheric_modeling}.

\section{Atmospheric modeling} \label{sec:atmospheric_modeling}
We employed both forward modeling and retrieval modeling techniques to interpret our
combined transmission spectrum. Initial retrieval analysis with our custom code
\texttt{exoretreivals} \citep{bixel2019, espinoza2019, weaver2020, mcgruder2020}
favored
models including stellar heterogeneity, but with the caveat that the corresponding spot
covering fractions needed to take up nearly a third of the star's surface area. From our
analysis of the ASAS-SN photometric light curve of HAT-P-23, shown in section 5, we
estimate an average spot coverage fraction of about 2.2\% to 2.6\% of the surface of the
star, so the result of the retrievals seems unphysical. We next turned to the
one-dimensional radiative-convective-chemical equilibrium model \texttt{ATMO}
\citep{goyal2018} to examine our spectrum, omitting contributions from stellar activity.
This analysis (described in Section~\ref{ssec:atmo}) favored a transmission spectrum
containing TiO/VO, which then informed our next step.

Given that TiO/VO can originate from spots with temperatures similar to M type stars, we
returned to retrievals to investigate if this signal persisted after taking stellar
activity into account. To address the previous problem with the retrievals (i.e. fitting
freely spot contamination introduced too many degrees of freedom, yielding to best fits
with unphysical spot properties), we assumed spot models instead, with the spot parameter
values derived in Section~\ref{sec:stell_act}.

Assuming that this signal in the transmission spectrum comes from spots on HAT-P-23, we
bound our retrievals to spot temperatures where TiO is stable in sun-like stars
$(T_\text{sp,lower}=\SI{2200}{K}, T_\text{sp,upper}=\SI{3800}{K})$, setting the lower and
upper bound on the corresponding spot covering fractions estimated in
Equation~\ref{eqn:covering_fraction} to $(f_\text{sp,lower} = 0.022,
f_\text{sp,upper}=0.026)$, respectively. We discuss the results of making this assumption
in Section~\ref{ssec:retrieval}.

\subsection{Forward modeling analysis} \label{ssec:atmo}
We used a generic grid of forward model transmission spectra as detailed in \citet{goyal2019} to interpret the observations of \hpb{}. This grid of models has been generated using the 1D-2D radiative-convective equilibrium model \texttt{ATMO} \citep{tremblin2015, amundsen2014, drummond2016, goyal2018}. The model transmission spectra in the grid are generated for a Jupiter"-like (radius and mass) planet around a Sun"-like star. However, this can be scaled to any hot Jupiter exoplanet with H$_2$-He dominated atmosphere using a scaling relationship detailed in \citet{goyal2019}. Each model in the grid assumes chemical equilibrium abundances and isothermal pressure-temperature (P-T) profiles. The model spectra includes H$_2$-H$_2$, H$_2$-He collision-induced absorption (CIA) and opacities due to H$_2$O, CO$_2$, CO, CH$_4$, NH$_3$, Na, K, Li, Rb, Cs, TiO, VO, FeH, PH$_3$, H$_2$S, HCN, SO$_2$ and C$_2$H$_2$. The entire grid spans 24 equilibrium temperatures from \SIrange{300}{2600}{K} in steps of \SI{100}{K}, six metallicities (\SIrange{0.1}{200}{}x solar), four planetary gravities (\SIrange{5}{50}{m/s^2}), four C/O ratios (\SIrange{0.35}{1.0}), two condensation regimes (local and rainout condensation) and four parameters each describing scattering hazes and uniform clouds. The haze parameter is defined as the multiplicative factor to the wavelength dependent Rayleigh scattering due to small particles, whilst the cloud parameter is defined as the multiplicative factor to uniform grey scattering, simulating the effects of a cloud deck, where a factor of 0 indicates no clouds, while a factor of 1 indicates extremely cloudy. Further details can be found in \citet{goyal2019}. % cloud opacity across all wavelengths.

\subsubsection{Forward model results and interpretation}
\begin{figure*}[htbp]
    \centering
    \includegraphics[width=0.98\linewidth]{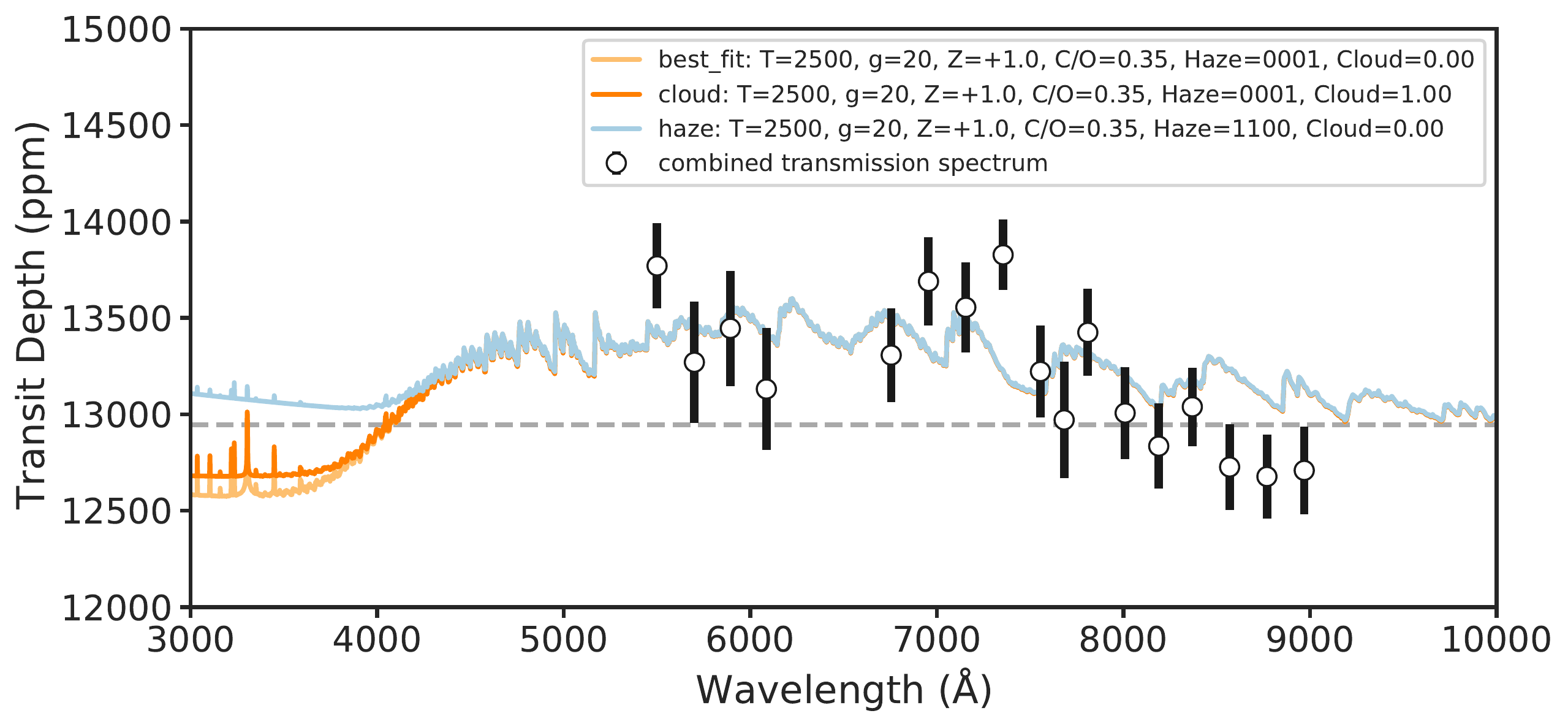}
    \caption{%
        Final combined transmission spectrum from Figure~\ref{fig:tspec_full} (filled
        white circles) compared to a best"-fit (yellow), cloud (orange), and haze (blue) model found from forward modeling on a
        generic grid. The weighted mean white-light curve depth is also
        shown here (grey dashed line) for comparison. We share the $\chi^2$ map of the entire grid in Figure~\ref{fig:tspec_forward_model_corner} of the Appendix.
        \pylink{https://icweaver.github.io/HAT-P-23b/notebooks/tspec_forward_model.html}
    }
    \label{fig:tspec_forward_model}
\end{figure*}

The best"-fit forward model is found by scaling each model in the grid to the parameters of HAT-P-23b and finding the model with the lowest $\chi^2$ throughout the entire grid. Figure \ref{fig:tspec_forward_model} shows this best"-fit forward model spectrum with the $\chi^2$ value of 34.25 and reduced $\chi^2$ value of 2.14 with 16 degrees of freedom (17 data points minus 1). We also note that while fitting the forward model grid to the observed spectrum, a wavelength"-independent constant vertical offset (transit depth) between model and observed spectra is a free parameter. The best"-fit model is consistent with a temperature of 2500\,K, ten times solar metallicity, and C/O ratio of 0.35, without any haze or clouds. For both the condensation regimes (local and rainout), we find the best"-fit model is the same, which is driven by the very high best"-fit temperature (2500\,K), where rarely any condensation occurs. Figure \ref{fig:tspec_forward_model_corner} shows the $\chi^2$ map of all the models in the grid when fitted to observations, along with contours of confidence intervals. There is a clear preference for high"-temperature models. There is also a preference for low C/O ratio models within one sigma, due to lack of spectral features of any carbon"-based species. However, the other parameters are less constrained. The rise in the transit depth that begins after \SI{4000}{\angstrom} in this model spectrum is due to TiO/VO opacity. Thus the best"-fit forward model for HAT-P-23b hints towards the presence of TiO/VO in its atmospheric limb. 

\subsection{Retrieval analysis} \label{ssec:retrieval}
With this key insight from our forward modeling analysis, we next performed a retrieval analysis with \texttt{exoretrievals} similar to the one described
in \citet{weaver2020}, with the key difference being now holding the spot parameters fixed
to values informed from the host star's photometry. We explored a suite of models for the
bounding spot parameters $(T_\text{sp,lower}, f_\text{sp,lower}) = (\SI{2200}{K}, 0.022)$
and $(T_\text{sp,upper}, f_\text{sp,upper}) = (\SI{3800}{K}, 0.026)$, and atmospheric
models including \Na, \K, and TiO, and assuming: i) a clear atmosphere (clear), ii) an
atmosphere with clouds (clear+cloud), iii) an atmosphere with hazes (clear+haze), and a
clear atmosphere in the presence of a starspot (clear+spot).

We ran these various models both fitting for the mean reference radius, $R_0$, and leaving
it fixed to our weighted mean white"-light value (using the $R_p/R_s$ values in
Table~\ref{tab:detrended_wlcs}), to investigate its degeneracy with clouds. All models
used the updated value for the stellar radius $(R_s=\SI{1.152}{\Rs})$ from
\citet{stassun2019}. In total, we explored 224 possible retrievals: 8 combinations of the
model types (clear, cloud, haze, spot) $\times$ 7 combinations of the three species
$\times$ 4 combinations of the fixed spot temperatures/covering fractions with and without
fitting for $R_0$.

\subsection{Retrieval results and interpretations}
\begin{figure*}[htbp]
    \centering
    \includegraphics[width=0.98\linewidth]{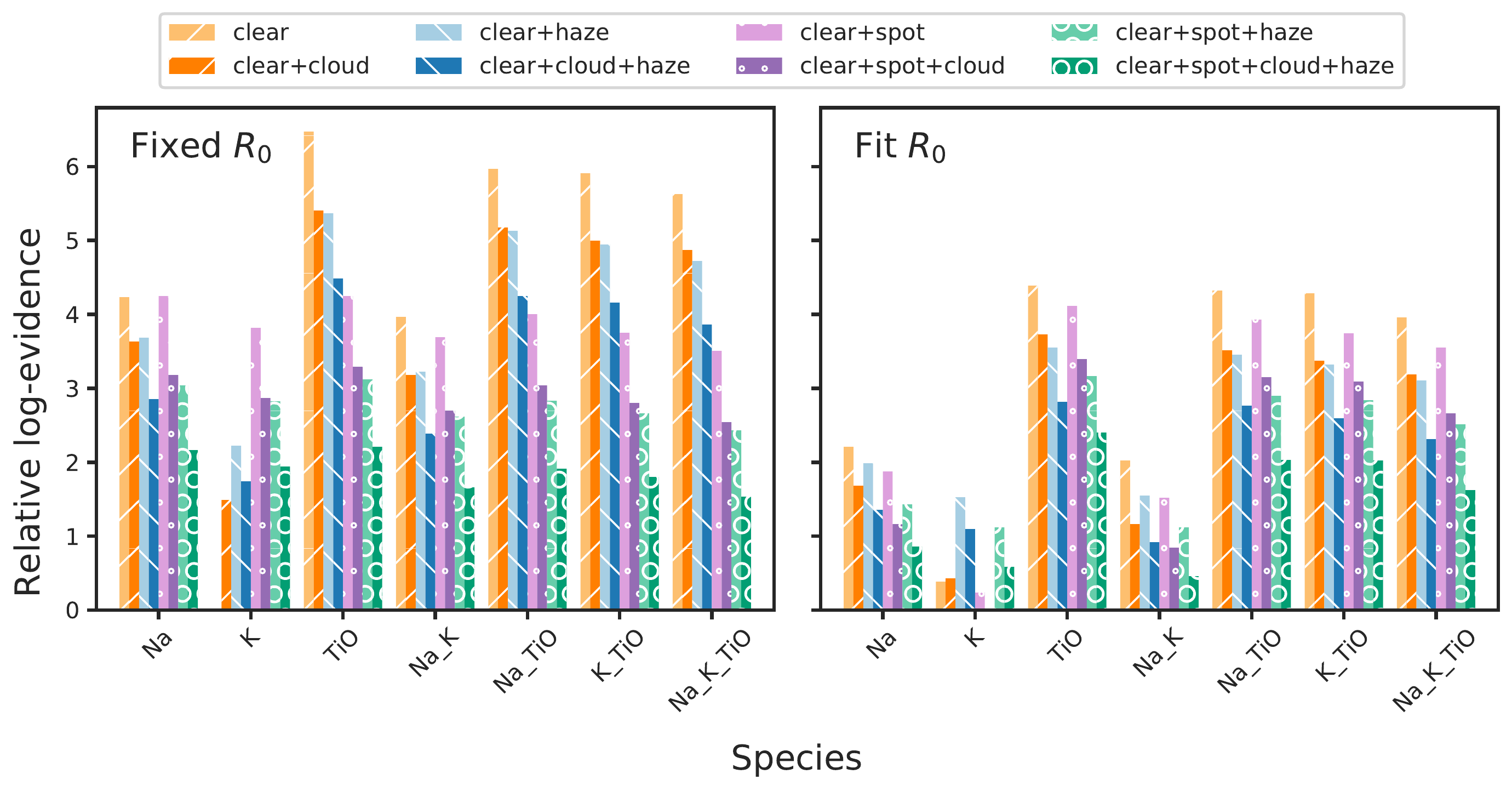}
    \caption{%
        Representative summary of Bayesian log-evidences of retrieved models relative to
        the model with the lowest log-evidence (left: Fixed $R_0$, K, clear; right: Fit
        $R_0$, K, clear+spot+cloud). Both sets correspond to $(T_\text{sp,lower},
        f_\text{sp,lower}) = (\SI{2200}{K}, 0.022)$, with $R_0 = \text{\MeanWLCDepth}$ fixed to
        our observed mean white"-light curve depth in the left panel. A trend can be seen
        favoring models with TiO present in the planet's atmosphere over contributions
        originating from stellar activity.
        \pylink{https://icweaver.github.io/HAT-P-23b/notebooks/retrieval_evidence.html}
    }
    \label{fig:retrieval_evidence}
\end{figure*}

\begin{figure*}[htbp]
    \centering
    \includegraphics[width=0.98\linewidth]{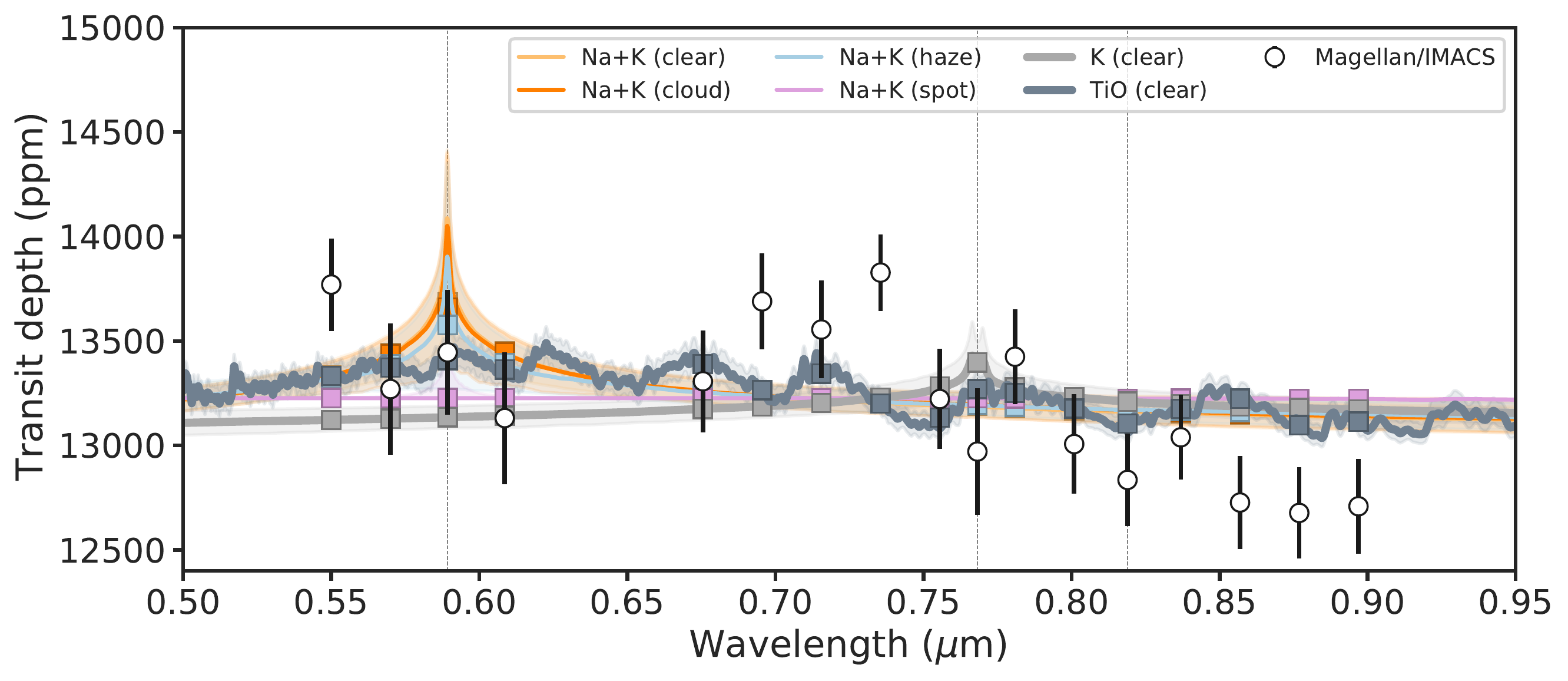}
    \caption{%
        Sample retrieved transmission spectra demonstrating the impact of different
        retrieval models on the overall shape of the spectrum when $R_0$ is held fixed
        (Figure~\ref{fig:retrieval_evidence}, left panel). We also include the models with
        the lowest (K, clear) and highest (TiO, clear) log-evidence in bold for
        comparison. Other models including TiO were omitted because of their visual
        similarity. We included the associated retrieved parameters for each model in
        Table~\ref{tab:retrieval_summary} of the Appendix.
        \pylink{https://icweaver.github.io/HAT-P-23b/notebooks/retrieval_tspec.html}
    }
    \label{fig:retrieval_tspec}
\end{figure*}

The suite of retrieval models described above all returned consistent results marginally
favoring TiO being present in the atmosphere of HAT-P-23b, with the most favored model corresponding to a sub-solar to super-solar TiO abundance of $10^{{-7.1 \pm 4}}$. We share a representative summary of
the Bayesian log-evidence of each model relative to the model with the lowest log-evidence
in Figure~\ref{fig:retrieval_evidence} and associated priors in
Table~\ref{tab:retrieval_priors}. We also show a selection of the retrieved transmission
spectra in Figure~\ref{fig:retrieval_tspec}, and the summary statistics for the
corresponding posterior distributions in Table~\ref{tab:retrieval_summary} of the
Appendix.

Although models including TiO in the atmosphere of the planet without
contributions from stellar activity tended to be preferred in all retrieval combinations
examined, the preference over stellar activity was not as large when fitting for $R_0$
(Figure~\ref{fig:retrieval_evidence}, right panel). For that case, the log-evidence for
every model was below 5, making the differences between specific models less statistically
significant. New data collected in the UV and IR would allow for more robust retrieval
results in this case. Finally, we note that because these retrievals are a 1D analysis,
our retrieved terminator planet temperatures are consistently cooler than the expected
equilibrium temperature of HAT-P-23b, as predicted by \citet{macdonald2020, pluriel2020}.

\begin{deluxetable*}{LlL}[htbp]
    \tablecaption{%
        Priors used to produce Figures~\ref{fig:retrieval_evidence} and
        \ref{fig:retrieval_tspec}.
        \label{tab:retrieval_priors}
    }
    \tablehead{
        \colhead{Parameter} & \colhead{Definition} & \colhead{Prior (uniform)}
    }
    \startdata
    f                    & reference planet radius normalization ($f R_0$) & [0.8, 1.2]   \\
    \log P_0                 & log10 cloud top pressure (bar)              & [-6, 6]      \\
    \Tp                      & planet atmospheric temperature (K)          & [100, 3000]  \\
    \log(\text{species})     & trace species mixing ratio                  & [-30, 0]     \\
    \log\sigma_\text{cloud}  & log10 grey opacity cross section            & [-50, 0]     \\
    \log a                   & log10 haze amplitude                        & [-30, 30]    \\
    \gamma_\text{haze}       & haze power-law                              & [-4, 0]      \\
    \Tchord                  & occulted stellar temperature (K)            & [5420, 6420] \\
    \Thet                    & star spot temperature (K)                   & [2200]       \\
    \fhet                    & spot covering fraction                      & [0.022]
    \enddata
\end{deluxetable*}

\subsection{Implications and broader context}
\begin{figure*}[htbp]
    \centering
    \includegraphics[width=0.98\linewidth]{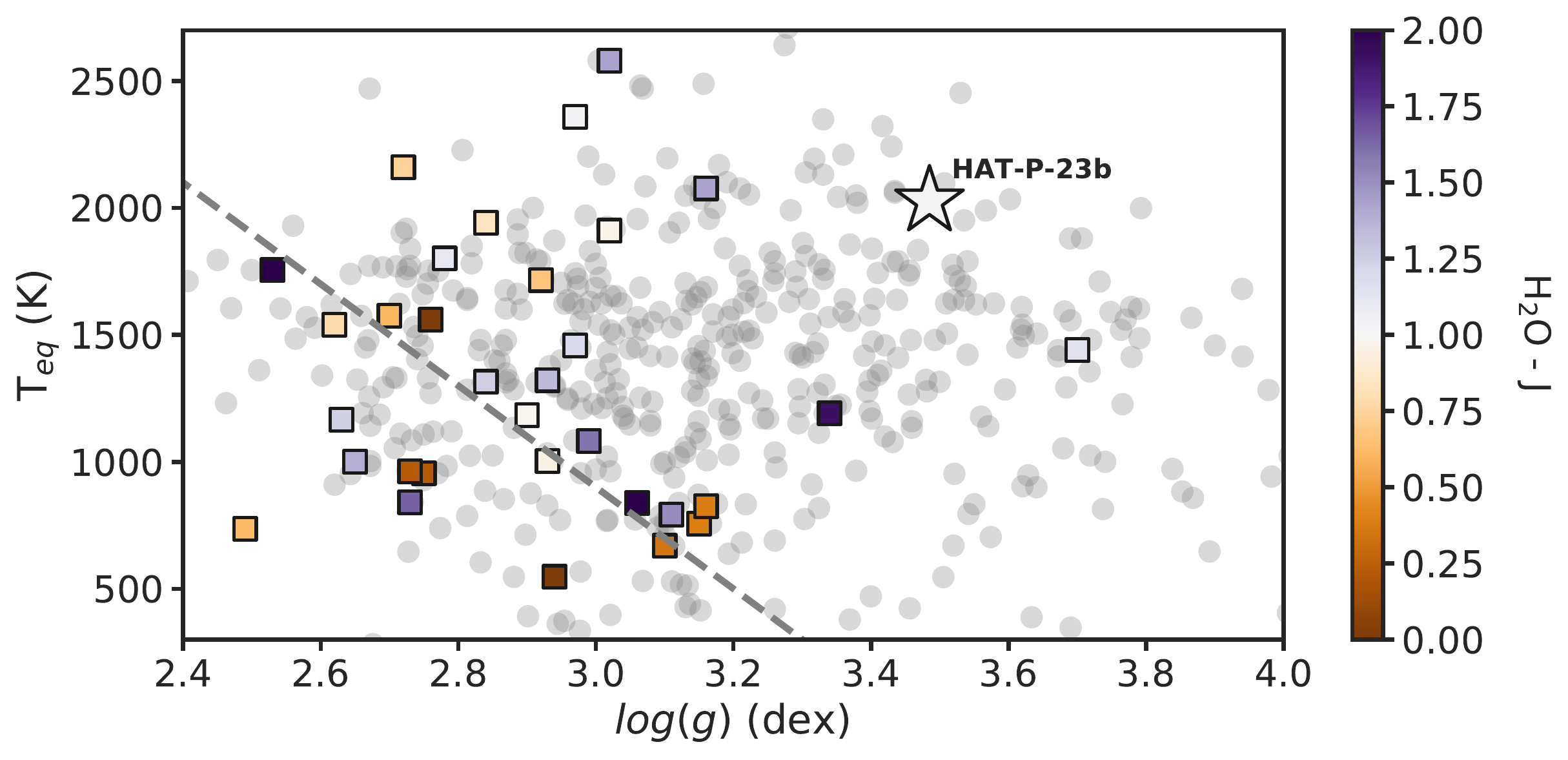}
    \caption{%
        Measured surface gravity vs. equilibrium temperature of re-analyzed exoplanets
        (colored squares) from Figure 13 of \citet{alam2020_1}, color coded by water
        feature strength. Exoplanets with mass and radius measurements are shown for
        comparison in grey. We overplot HAT-P-23b for comparison to show its unique
        position in this parameter space and motivate future space-based missions to
        measure its \SI{1.4}{\micron} \water bandhead amplitude, which will provide a good
        test for the delineation between clear and cloudy exoplanets based on water
        feature strength proposed by \citet{stevenson2016} (gray dashed line).
        \pylink{https://icweaver.github.io/HAT-P-23b/notebooks/cloudy_vs_clear.html}
    }
    \label{fig:cloudy_vs_clear}
\end{figure*}

The trend between exoplanets with known surface gravity and equilibrium temperature vs.
water feature strength in the near-IR, proposed by \citet{stevenson2016}, would predict that
HAT-P-23b has a clear atmosphere. In the re-analysis of this relationship by
\citet{alam2020_1} using new data from the Hubble Space Telescope's Wide Field Camera 3
(\textit{HST/WFC3}) \citep{wakeford2019}, they note that this divide may not be as
distinct as initially suggested. Because of its high surface gravity $(g \approx
\SI{30}{m/s^2})$ and relatively high equilibrium temperature $(T_{eq} = \SI{2027}{K})$,
HAT-P-23b tests a region of parameter space previously unexplored (see
Figure~\ref{fig:cloudy_vs_clear}). Measurements of its transmission spectrum in the
$J$-band (\SIrange{1.36}{1.44}{\micron}) and baseline region
(\SIrange{1.22}{1.30}{\micron}) to determine the planet's water feature signal would
provide insight into this potential trend in exoplanet atmosphere types. For example, a
single transit in the \SIrange{7500}{18000}{\angstrom} IR range with G141 grism aboard
\textit{HST/WFC3} would provide the novel data needed for such an analysis. In addition, a
second transit in the UV/Visible channel with the \SIrange{2000}{8000}{\angstrom} grism
(G280) \citep{wakeford2020} would provide the wide wavelength coverage necessary to
perform a robust retrieval of the combined transmission spectrum and test our preliminary
result.

\section{Summary and conclusions} \label{sec:conclusion}
We collected, extracted, and combined transmission spectra of HAT-P-23b from
\textit{Magellan/IMACS} from five transit events between 2016 and 2018, spanning a total
wavelength coverage of \SIrange{5200}{9269}{\angstrom}. We summarize our findings from
this combined transmission spectrum below:
\begin{itemize}
    \item The average precision we achieve per wavelength bin (\SI{247}{ppm})
          is comparable to the predicted 5"-scale"-height signal assuming a hydrogen-dominated atmosphere (\SI{384}{ppm}). We note that the actual signal
          could be appreciably smaller, requiring additional transits to build
          sufficient signal to make any strong claims about the planet's
          atmosphere.
    \item We perform a generic forward modeling analysis with \texttt{ATMO} to
          begin this characterization. We find that the best fitting model favors
          a high equilibrium temperature ($\Teq \approx \SI{2500}{K}$), which we
          note may be driven by potential TiO/VO features present in the
          transmission spectrum.
    \item We perform a comprehensive retrieval analysis exploring all combinations of
          atomic/molecular species and atmosphere types tested. Informed by photometric activity data, we find a consistent
          trend marginally favoring the presence of TiO in the atmosphere of HAT-P-23b, but
          note that this trend is not as strong when fitting for the mean transit
          depth.
    \item We place HAT-P-23b in the context of other exoplanets that
          have been analyzed in terms of their measured surface gravities and
          equilibrium temperatures in the search for potential predictive trends
          in the cloudiness vs. clarity of their atmospheres. With our current results
          favoring a clear atmosphere with TiO for HAT-P-23b, the planet falls in
          the predicted location for clear planets by \citet{stevenson2016}.
    \item Finally, we recommend that because this suggested trend is based on measurements
          made in the near-IR, this unique target (due to its large surface gravity)
          should be followed-up with HST to provide this necessary data and provide
          a good test for this potential trend.
\end{itemize}

\acknowledgments
We thank the anonymous referee and data editor for helpful comments and feedback. This paper includes data gathered with the 6.5 meter Magellan Telescopes located at Las
Campanas Observatory, Chile. We thank Jennifer Fienco, the Las Campanas Observatory staff, and the observing personnel for providing the facilities and guidance necessary for making the collection
of this high"-quality dataset possible. This research has made use of the VizieR catalogue
access tool, CDS, Strasbourg, France \citep{asassn2000}. A.J.\ acknowledges support from
ANID – Millennium Science Initiative – ICN12\_009. The results reported herein benefited
from collaborations and/or in- formation exchange within NASA's Nexus for Exoplanet System
Science (NExSS) research coordination network sponsored by NASA's Science Mission
Directorate, and funding through the NExSS Earths in Other Solar System (PI: Apai) and
ACCESS (PI: L\'opez-Morales) teams. ICW and MLM thank the Brinson Foundation for
supporting this project.
B.V.R. thanks the Heising-Simons Foundation for support.
This research used computing resources from the Smithsonian
Institution High Performance Cluster (SI/HPC).

\software{%
    Astropy \citep{astropy:2013, astropy:2018},
    batman \citep{batman},
    george \citep{george},
    multinest \citep{multinest2008, multinest2009, multinest2019},
    NumPy \citep{numpy},
    SciPy \citep{scipy},
    corner \citep{corner},
    jupyter \citep{jupyter},
    pandas \citep{pandas},
    matplotlib \citep{matplotlib},
    seaborn \citep{seaborn},
    Measurements.jl \citep{measurements.jl}
}


\bibliographystyle{yahapj}
\bibliography{paper}
%\bibliography{sample63}{}
%\bibliographystyle{aasjournal}

%% the collaboration and author truncation commands are used. It has to


\appendix

Here we provide additional figures and tables referenced in the main text.

\begin{figure*}[htbp]
\gridline{
    \fig{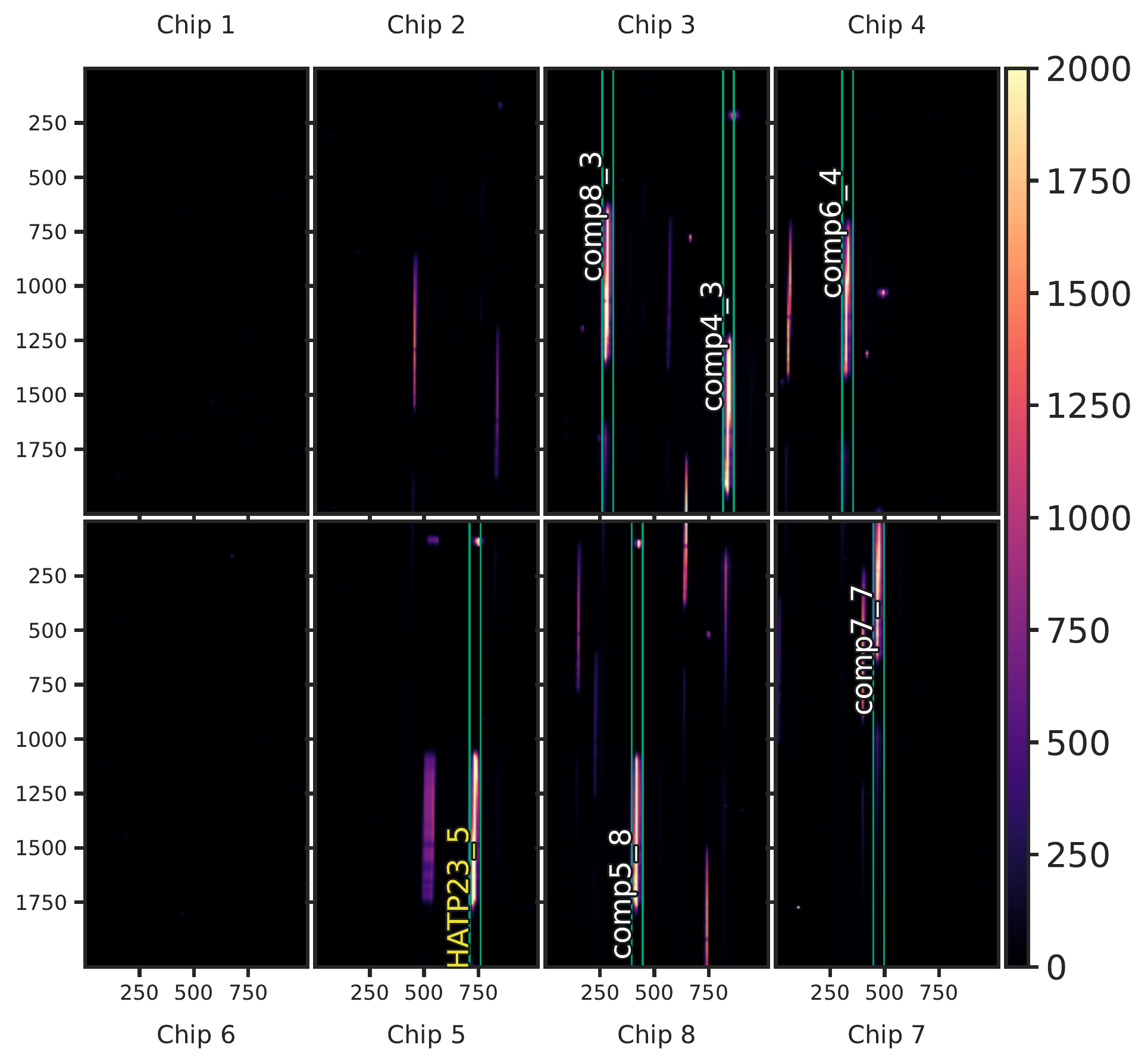}{0.3\linewidth}
        {(a) Transit 1}
    \fig{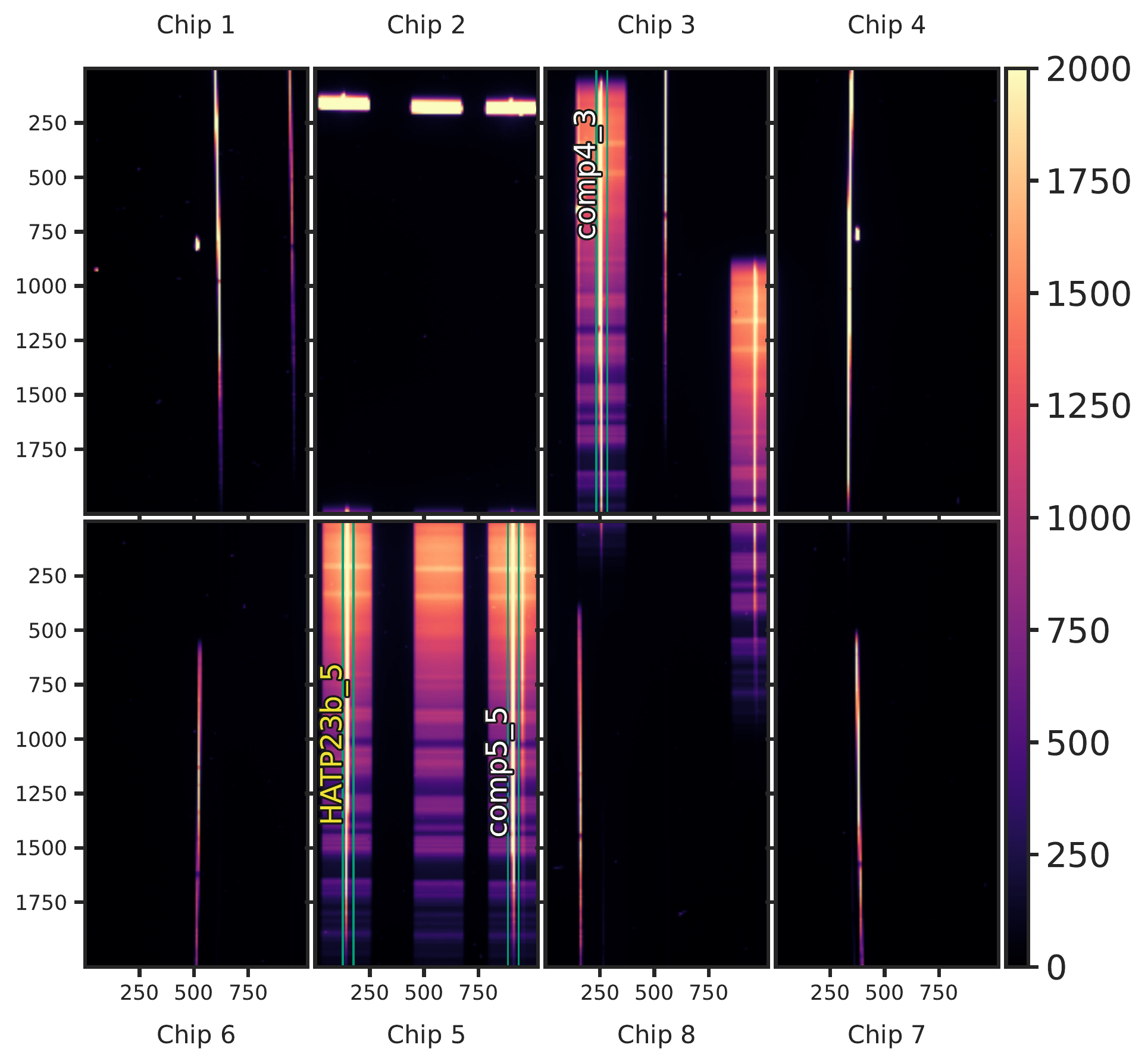}{0.3\linewidth}
        {(b) Transit 2}
    \fig{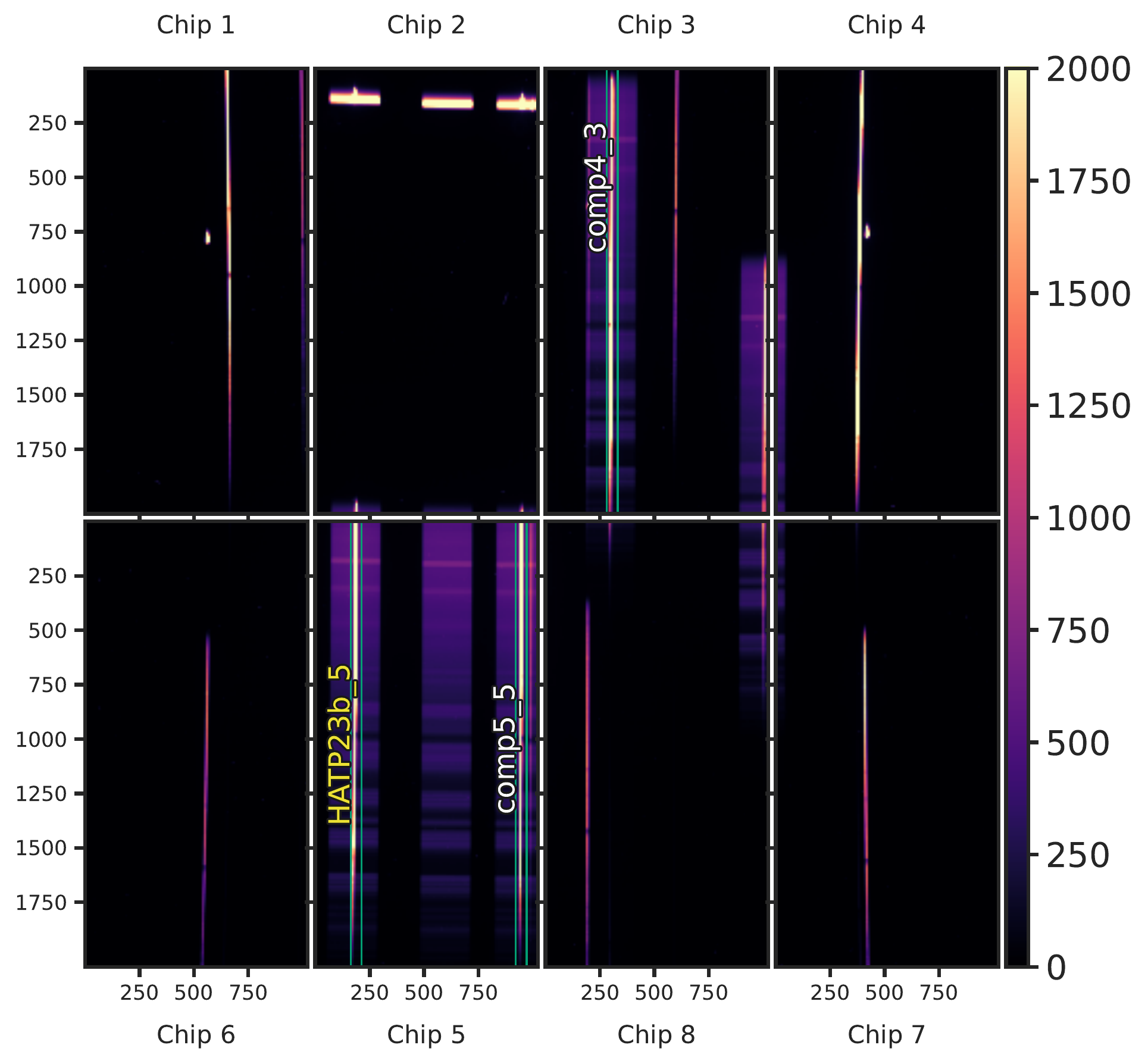}{0.3\linewidth}
        {(c) Transit 3}
}
\gridline{
    \fig{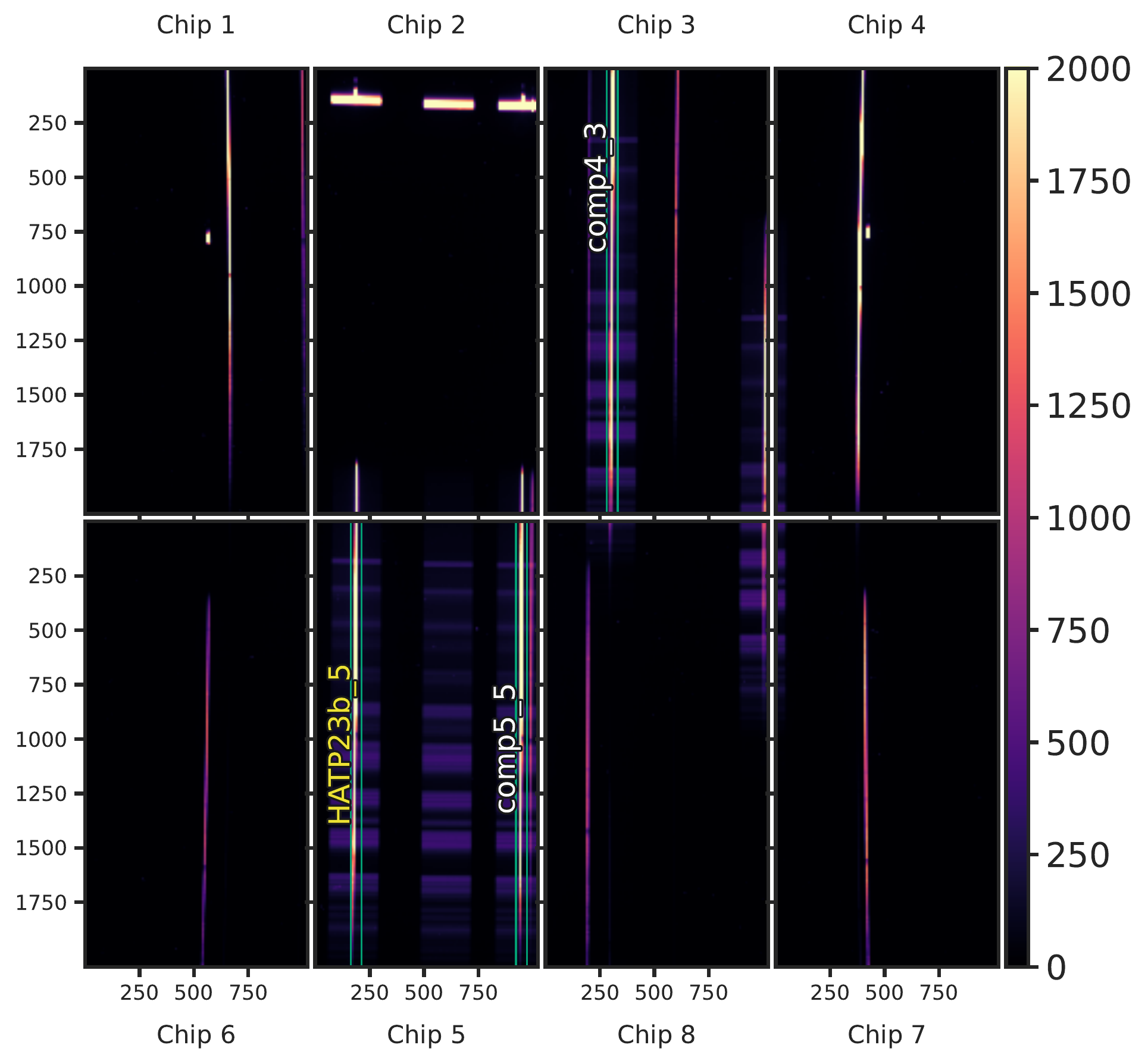}{0.3\linewidth}
        {(d) Transit 4}
    \fig{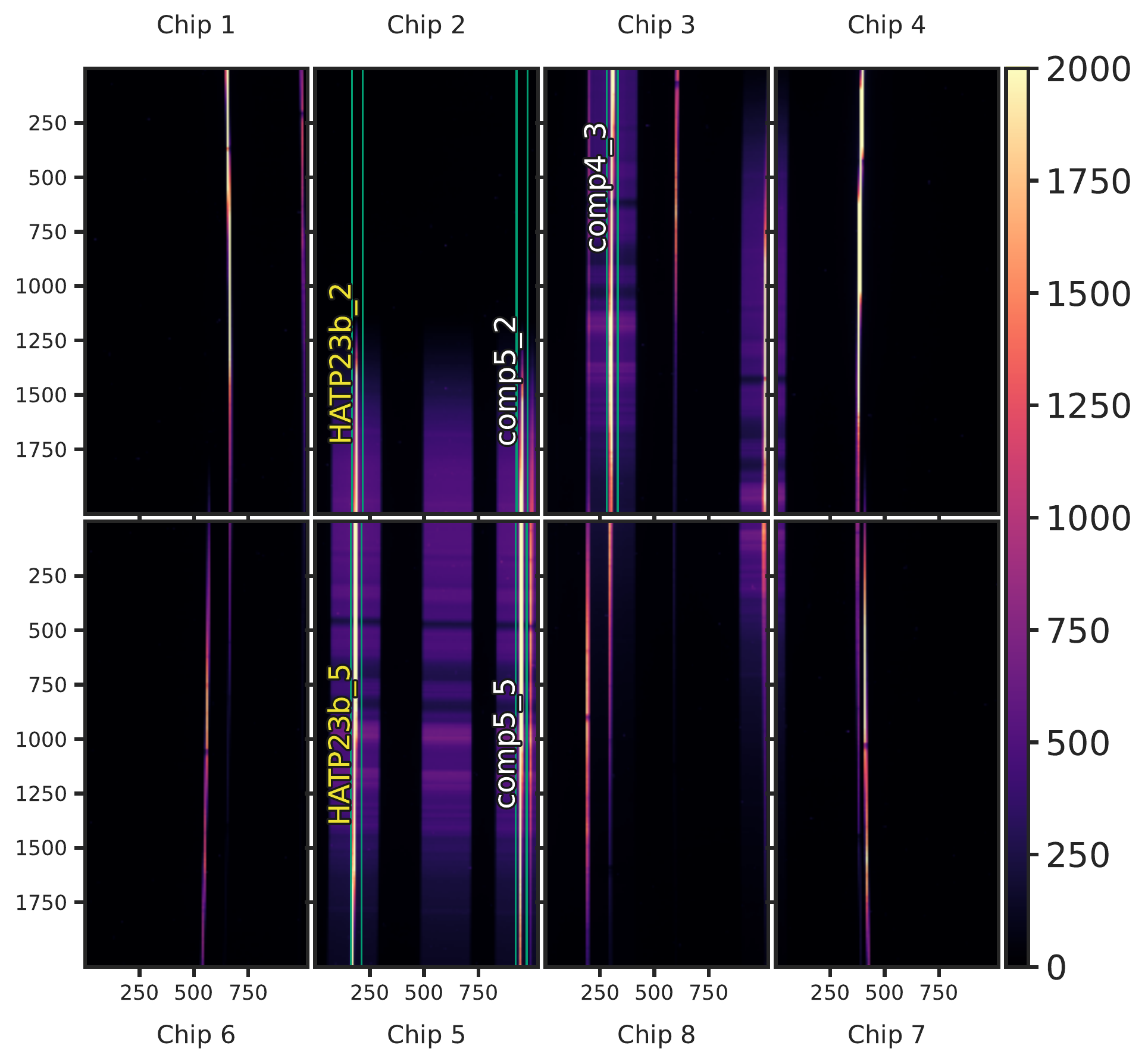}{0.3\linewidth}
        {(e) Transit 5}
}
\caption{%
        Sample science frames of 2D spectra collected from all nights on
        \textit{Magellan/IMACS} used in our final analysis, colored by counts. The presence of a
        contaminating source in the slit for comp5 set the maximum aperture size for all
        nights (show in green).
        \pylink{https://icweaver.github.io/HAT-P-23b/notebooks/raw_frames.html}
}
\label{fig:raw_frames}
\end{figure*}

\begin{figure*}[htbp]
    \centering
    \includegraphics[width=0.8\linewidth]{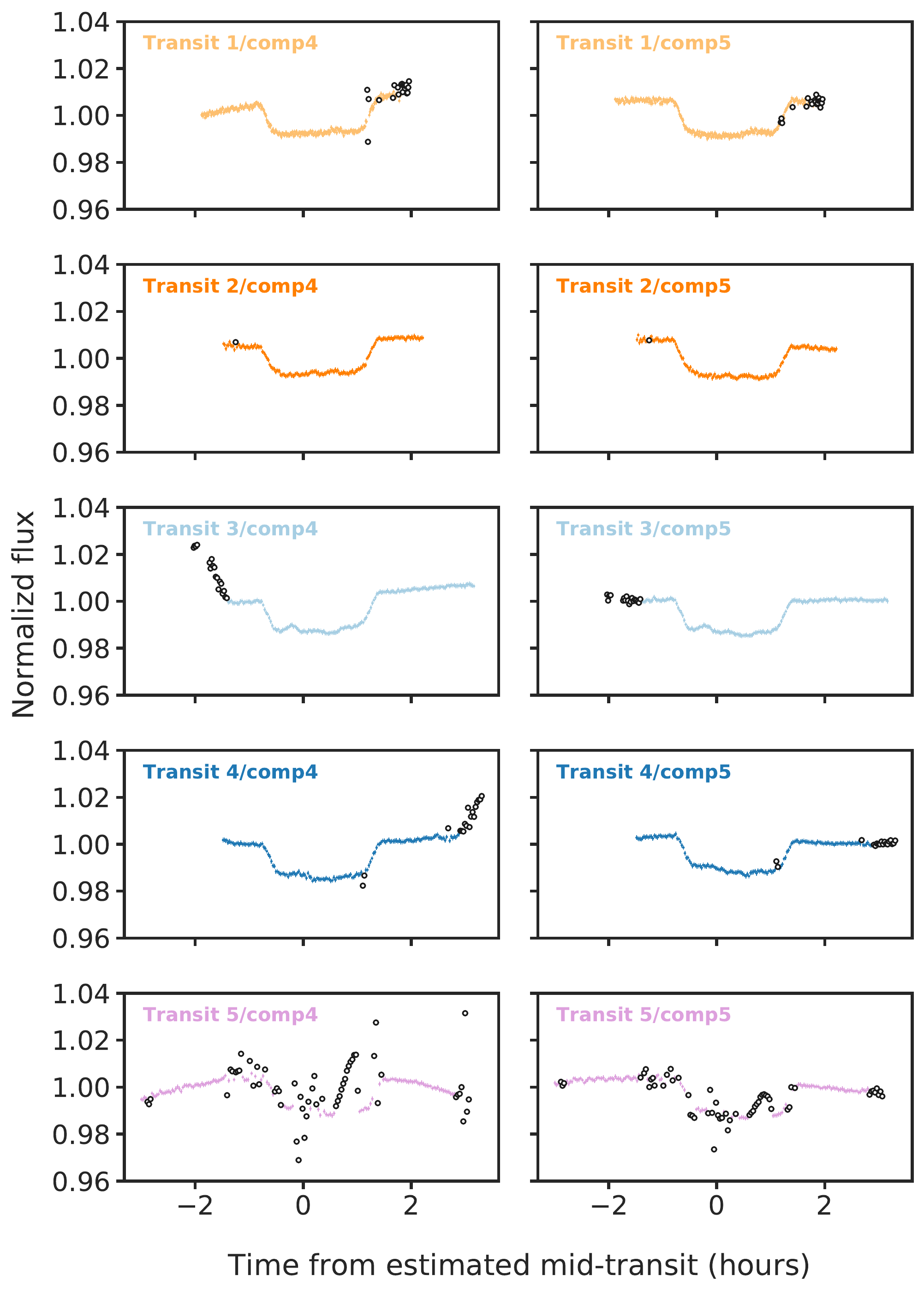}
    \caption{%
            Divided white"-light curves for each night used in our final analysis. The raw WLC
            for HAT-P-23 divided by the raw WLC for comp4 and comp5 is shown in the first and
            second column, respectively. The union of identified outliers for each night is
            highlighted in white circles.
            \pylink{https://icweaver.github.io/HAT-P-23b/notebooks/extracted_wlcs.html}
    }
    \label{fig:extracted_wlcs}
\end{figure*}

\begin{deluxetable}{cccc}[htbp]
    \tablecaption{%
        Wavelength bins (in \angstrom) used in main analysis.
        \label{tab:bins}
    }
    \tablehead{
        \colhead{Wav start} & \colhead{Wav end} &
        \colhead{Wav diff} & \colhead{Wav cen}
    }
    \startdata
    5200.0    & 5400.0  & 200.0    & 5300.0  \\
    5400.0    & 5600.0  & 200.0    & 5500.0  \\
    5600.0    & 5800.0  & 200.0    & 5700.0  \\
    5800.0    & 5985.8  & 185.8    & 5892.9  \\
    5985.8    & 6185.8  & 200.0    & 6085.8  \\
    6655.0    & 6855.0  & 200.0    & 6755.0  \\
    6855.0    & 7055.0  & 200.0    & 6955.0  \\
    7055.0    & 7255.0  & 200.0    & 7155.0  \\
    7255.0    & 7455.0  & 200.0    & 7355.0  \\
    7455.0    & 7655.0  & 200.0    & 7555.0  \\
    7655.0    & 7709.0  & 54.0     & 7682.0  \\
    7709.0    & 7909.0  & 200.0    & 7809.0  \\
    7909.0    & 8109.0  & 200.0    & 8009.0  \\
    8109.0    & 8269.0  & 160.0    & 8189.0  \\
    8269.0    & 8469.0  & 200.0    & 8369.0  \\
    8469.0    & 8669.0  & 200.0    & 8569.0  \\
    8669.0    & 8869.0  & 200.0    & 8769.0  \\
    8869.0    & 9069.0  & 200.0    & 8969.0  \\
    9069.0    & 9269.0  & 200.0    & 9169.0
    \enddata
\end{deluxetable}

\begin{deluxetable}{cccc}[htbp]
    \tablecaption{%
        Wavelength bins (in \angstrom) used in Figure~\ref{fig:tspec_species}.
        \label{tab:species}
    }
    \tablehead{
        \colhead{Wav start} & \colhead{Wav end} &
        \colhead{Wav diff} & \colhead{Wav cen}
    }
    \startdata
    5780.4    & 5825.4  & 45.0     & 5802.9  \\
    5825.4    & 5870.4  & 45.0     & 5847.9  \\
    5870.4    & 5915.4  & 45.0     & 5892.9  \\
    5915.4    & 5960.4  & 45.0     & 5937.9  \\
    5960.4    & 6005.4  & 45.0     & 5982.9  \\
    7657.0    & 7667.0  & 10.0     & 7662.0  \\
    7667.0    & 7677.0  & 10.0     & 7672.0  \\
    7677.0    & 7687.0  & 10.0     & 7682.0  \\
    7687.0    & 7697.0  & 10.0     & 7692.0  \\
    7697.0    & 7707.0  & 10.0     & 7702.0  \\
    8089.0    & 8129.0  & 40.0     & 8109.0  \\
    8129.0    & 8169.0  & 40.0     & 8149.0  \\
    8169.0    & 8209.0  & 40.0     & 8189.0  \\
    8209.0    & 8249.0  & 40.0     & 8229.0  \\
    8249.0    & 8289.0  & 40.0     & 8269.0
    \enddata
\end{deluxetable}

\begin{figure*}[htbp]
    \centering
    \includegraphics[width=0.8\linewidth]{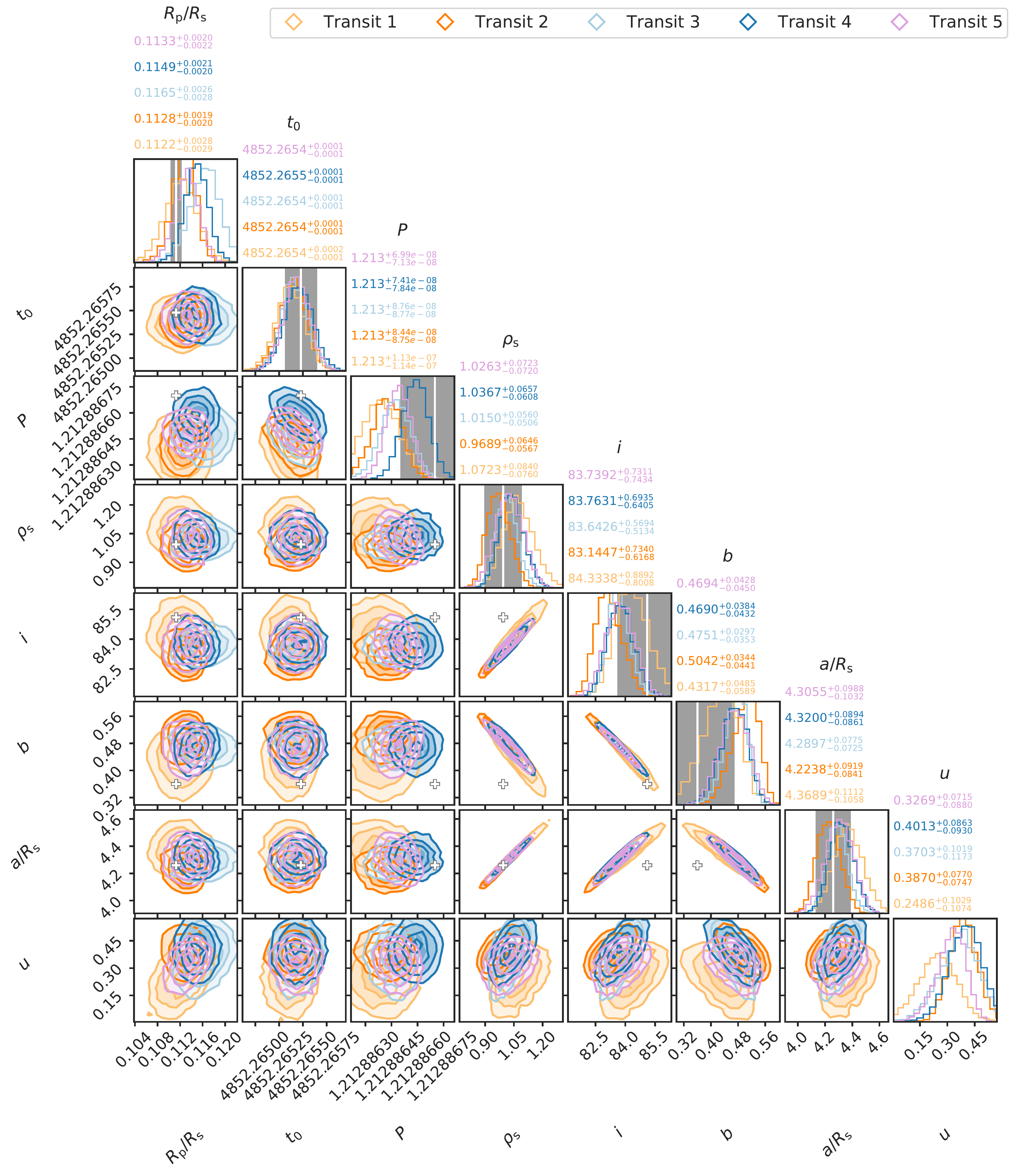}
    \caption{%
        Detrended white"-light curve corner plots for all nights used in our final
        analysis, color"-coded by transit. The reported values (highlighted, white crosses) are from a
        self"-consistent analysis (described in Section~\ref{sec:intro}) of the most
        recent transit and RV data available. The associated WLCs are presented in
        Figure~\ref{fig:detrended_wlcs} and tabulated values of our best"-fit parameters
        are shared in Table~\ref{tab:detrended_wlcs}.
        \pylink{https://icweaver.github.io/HAT-P-23b/notebooks/detrended_wlcs_corners.html}
    }
    \label{fig:detrended_wlcs_corners}
\end{figure*}

\begin{figure*}[htbp]
    \centering
    \includegraphics[width=0.8\linewidth]{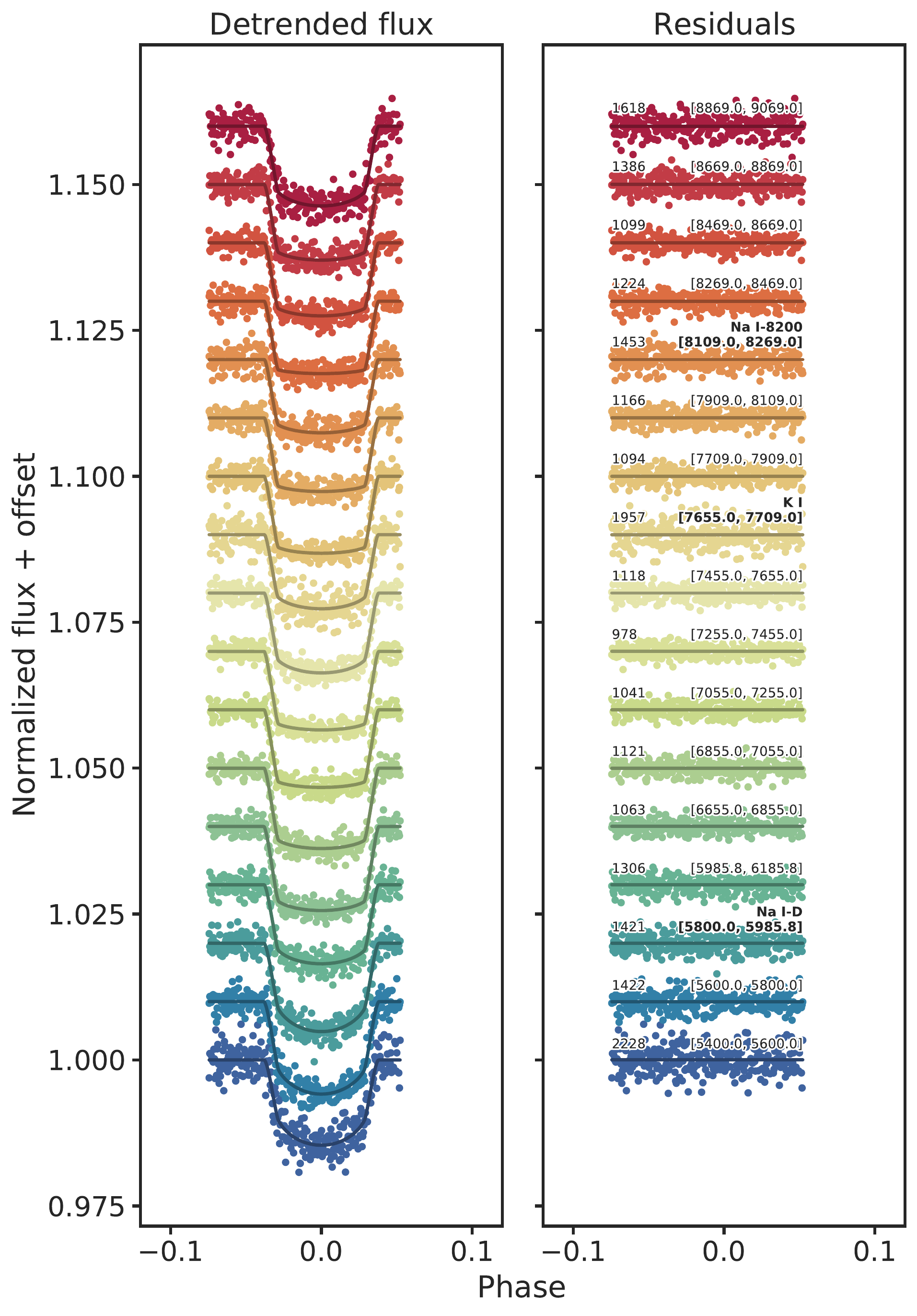}
    \caption{%
        Detrended binned light curves for Transit 1 used to build the final transmission
        spectrum in Figure~\ref{fig:tspec_full}. The standard deviation in the residuals
        and corresponding wavelength bin are annotated on the left and right side of the
        right"-hand panel, respectively. The tabulated data for these binned light curves
        are available in Table~\ref{tab:tspec_full}.
        \pylink{https://icweaver.github.io/HAT-P-23b/notebooks/detrended_binned_lcs.html}
    }
    \label{fig:detrended_binned_lcs_transit_1}
\end{figure*}

\begin{figure*}[htbp]
    \centering
    \includegraphics[width=0.8\linewidth]{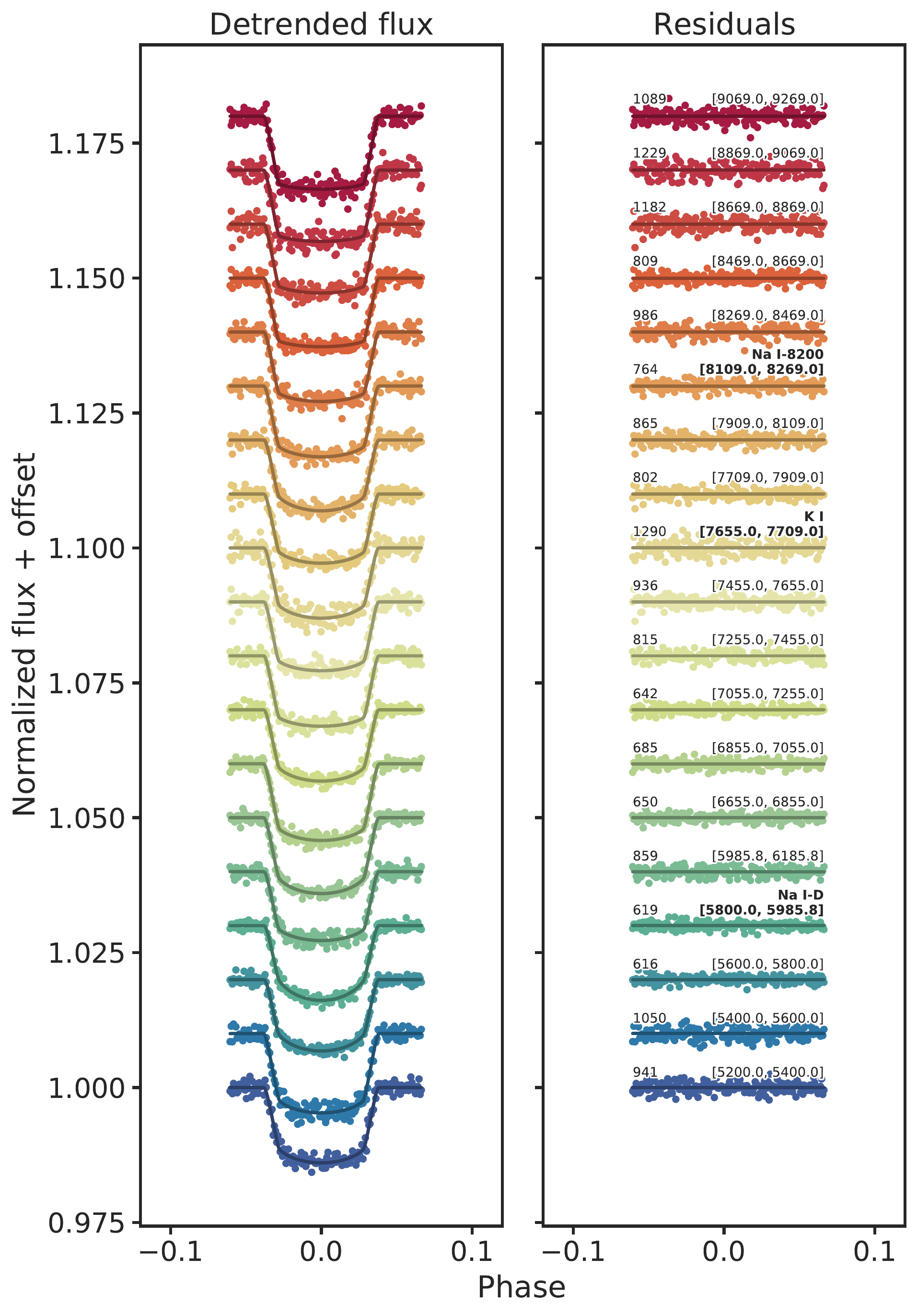}
    \caption{%
        Same as Figure~\ref{fig:detrended_binned_lcs_transit_1}, but for Transit 2.
        \pylink{https://icweaver.github.io/HAT-P-23b/notebooks/detrended_binned_lcs.html}
    }
    \label{fig:detrended_binned_lcs_transit_2}
\end{figure*}

\begin{figure*}[htbp]
    \centering
    \includegraphics[width=0.8\linewidth]{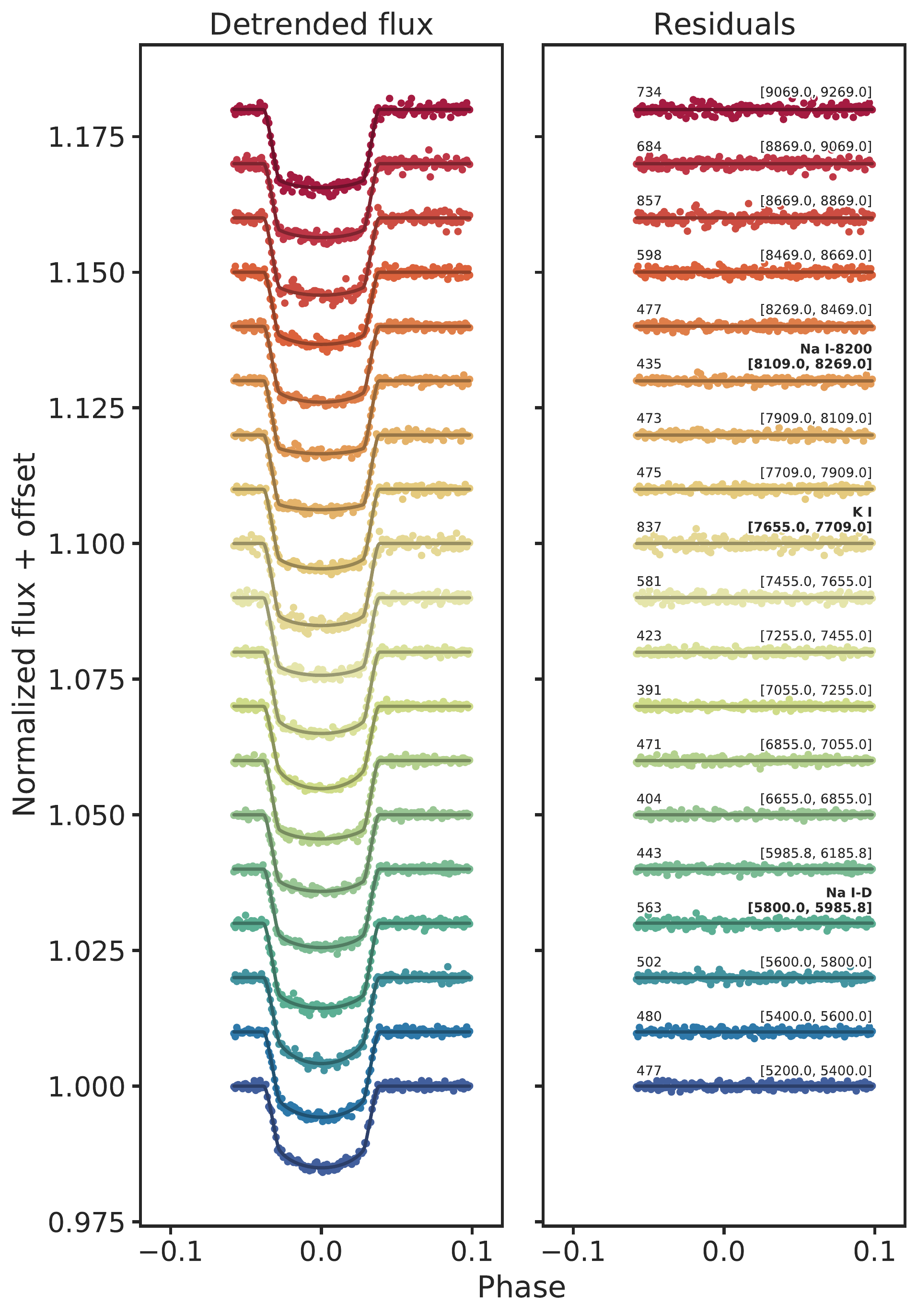}
    \caption{%
        Same as Figure~\ref{fig:detrended_binned_lcs_transit_1}, but for Transit 3.
        \pylink{https://icweaver.github.io/HAT-P-23b/notebooks/detrended_binned_lcs.html}
    }
    \label{fig:detrended_binned_lcs_transit_3}
\end{figure*}

\begin{figure*}[htbp]
    \centering
    \includegraphics[width=0.8\linewidth]{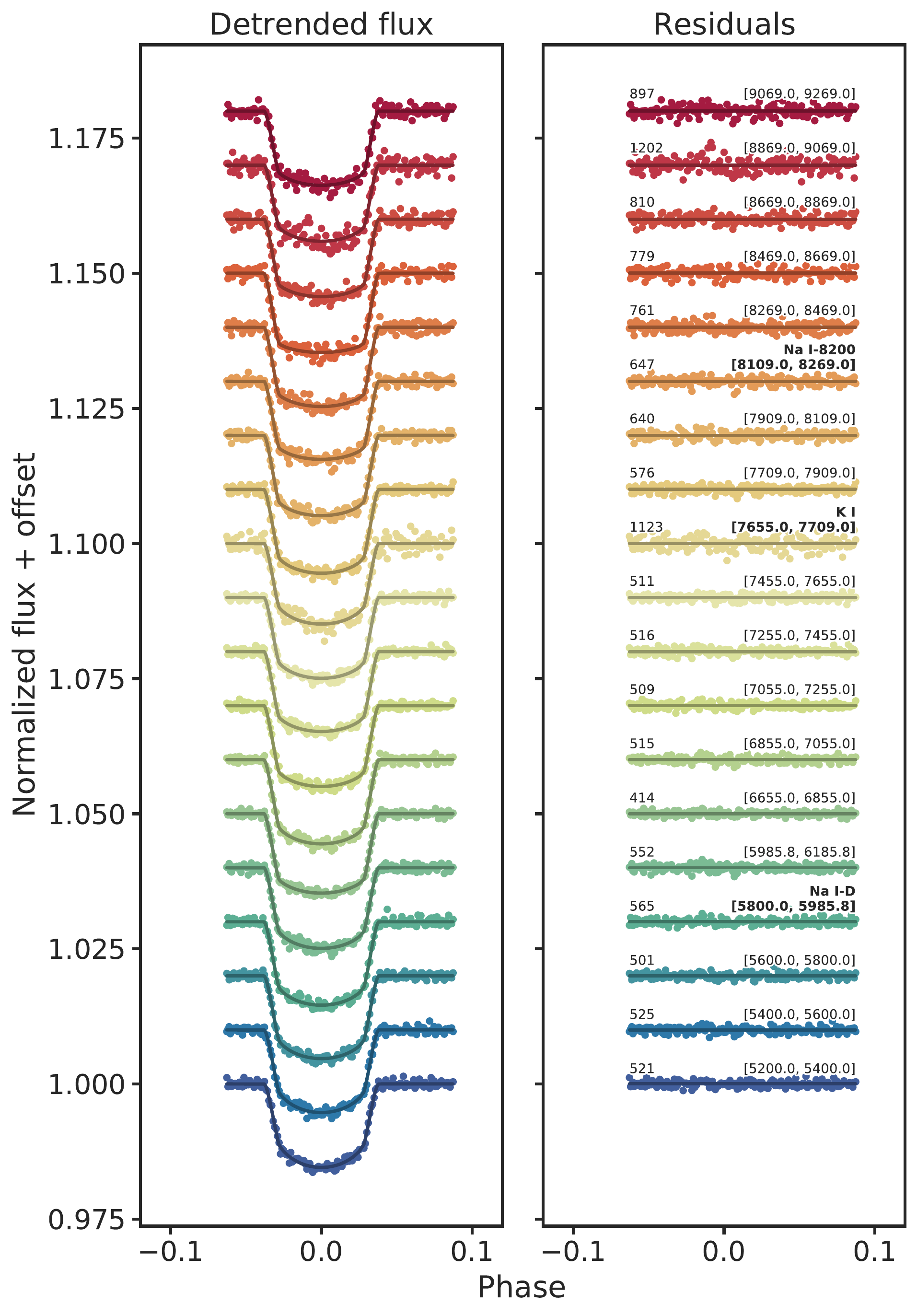}
    \caption{%
        Same as Figure~\ref{fig:detrended_binned_lcs_transit_1}, but for Transit 4.
        \pylink{https://icweaver.github.io/HAT-P-23b/notebooks/detrended_binned_lcs.html}
    }
    \label{fig:detrended_binned_lcs_transit_4}
\end{figure*}

\begin{figure*}[htbp]
    \centering
    \includegraphics[width=0.8\linewidth]{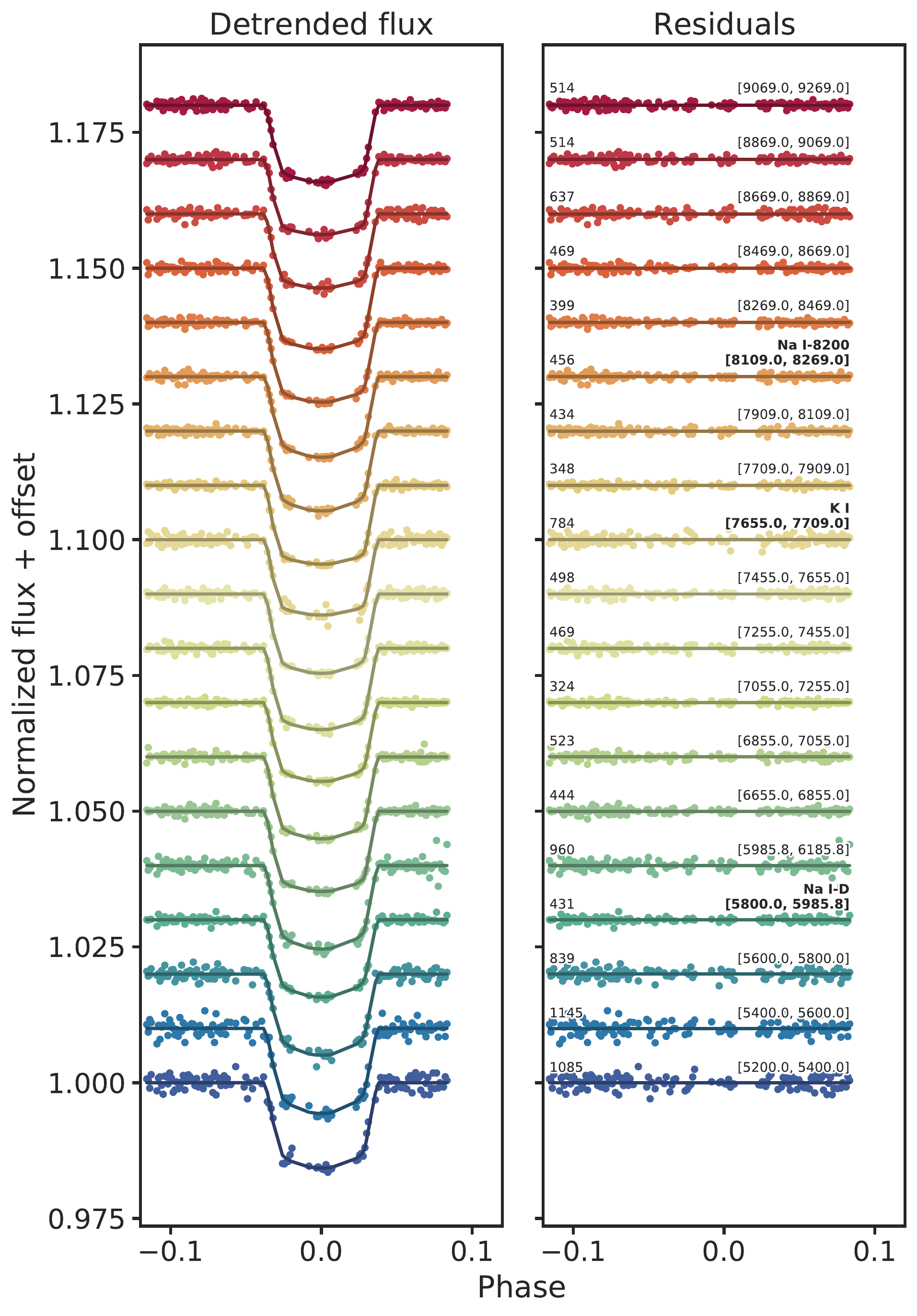}
    \caption{%
        Same as Figure~\ref{fig:detrended_binned_lcs_transit_1}, but for Transit 5.
        \pylink{https://icweaver.github.io/HAT-P-23b/notebooks/detrended_binned_lcs.html}
    }
    \label{fig:detrended_binned_lcs_transit_5}
\end{figure*}

%\begin{longrotatetable}
\begin{deluxetable*}{RRRRRRRR}
    %\tabletypesize{\scriptsize}
    \tablecaption{%
        Final offset, tabulated transit depths (in ppm) from Figure~\ref{fig:tspec_full}, corresponding
        to binned light curves in \mbox{Figures \ref{fig:detrended_binned_lcs_transit_1}
        -- \ref{fig:detrended_binned_lcs_transit_5}}. For reference, the weighted mean
        white"-light curve transit depth is \MeanWLCDepthErr.
        \label{tab:tspec_full}
    }
    \tablehead{
        \colhead{Wavelength (\angstrom)} &
        \colhead{Transit 1} &
        \colhead{Transit 2} &
        \colhead{Transit 3} &
        \colhead{Transit 4} &
        \colhead{Transit 5} &
        \colhead{Combined}
    }
    \startdata
    5400.0 - 5600.0 & 13100^{+1300}_{-1400} & 13900^{+300}_{-300}   & 13600^{+600}_{-600}   & 13300^{+600}_{-600} & 14000^{+600}_{-600} & 13800 \pm 200 \\
5600.0 - 5800.0 & 14200^{+1400}_{-1400} & 12000^{+1200}_{-1100} & 13300^{+700}_{-700}   & 13300^{+600}_{-600} & 13400^{+500}_{-500} & 13300 \pm 300 \\
5800.0 - 5985.8 & 13600^{+1100}_{-1200} & 11900^{+1100}_{-1000} & 14000^{+700}_{-800}   & 13700^{+400}_{-400} & 13000^{+800}_{-700} & 13500 \pm 300 \\
5985.8 - 6185.8 & 12800^{+1000}_{-1000} & 12300^{+1000}_{-500}  & 12800^{+1200}_{-1300} & 13200^{+500}_{-500} & 14000^{+700}_{-700} & 13200 \pm 300 \\
6655.0 - 6855.0 & 14000^{+900}_{-800}   & 13000^{+600}_{-700}   & 12600^{+500}_{-600}   & 13200^{+600}_{-600} & 13800^{+400}_{-400} & 13300 \pm 200 \\
6855.0 - 7055.0 & 13100^{+900}_{-900}   & 13400^{+400}_{-500}   & 13000^{+600}_{-600}   & 14000^{+600}_{-600} & 14100^{+400}_{-400} & 13700 \pm 200 \\
7055.0 - 7255.0 & 13000^{+700}_{-700}   & 12600^{+900}_{-900}   & 12900^{+700}_{-700}   & 13800^{+600}_{-600} & 13800^{+300}_{-300} & 13600 \pm 200 \\
7255.0 - 7455.0 & 13400^{+800}_{-700}   & 12500^{+600}_{-600}   & 13500^{+600}_{-600}   & 13300^{+500}_{-400} & 14300^{+200}_{-300} & 13800 \pm 200 \\
7455.0 - 7655.0 & 12900^{+1000}_{-1000} & 12200^{+600}_{-600}   & 12800^{+600}_{-600}   & 13400^{+400}_{-400} & 14000^{+500}_{-500} & 13200 \pm 200 \\
7655.0 - 7709.0 & 12200^{+900}_{-800}   & 12300^{+700}_{-700}   & 13800^{+1000}_{-1000} & 13500^{+700}_{-700} & 13200^{+500}_{-500} & 13000 \pm 300 \\
7709.0 - 7909.0 & 13100^{+800}_{-800}   & 12200^{+600}_{-600}   & 13000^{+500}_{-400}   & 14000^{+700}_{-700} & 14100^{+400}_{-400} & 13400 \pm 200 \\
7909.0 - 8109.0 & 12400^{+600}_{-600}   & 12200^{+700}_{-700}   & 12600^{+500}_{-500}   & 13600^{+500}_{-600} & 13600^{+400}_{-400} & 13000 \pm 200 \\
8109.0 - 8269.0 & 12500^{+400}_{-300}   & 12600^{+700}_{-800}   & 12300^{+500}_{-500}   & 13200^{+500}_{-500} & 13700^{+500}_{-500} & 12900 \pm 200 \\
8269.0 - 8469.0 & 12600^{+400}_{-400}   & 12600^{+800}_{-900}   & 12400^{+400}_{-400}   & 13400^{+500}_{-500} & 13800^{+400}_{-400} & 13100 \pm 200 \\
8469.0 - 8669.0 & 12300^{+600}_{-600}   & 12400^{+600}_{-500}   & 11800^{+500}_{-500}   & 13600^{+500}_{-500} & 13400^{+400}_{-400} & 12700 \pm 200 \\
8669.0 - 8869.0 & 12600^{+700}_{-600}   & 12300^{+400}_{-500}   & 12800^{+400}_{-400}   & 12900^{+400}_{-400} & 12800^{+700}_{-700} & 12700 \pm 200 \\
8869.0 - 9069.0 & 13100^{+800}_{-900}   & 12800^{+400}_{-500}   & 12300^{+600}_{-500}   & 12700^{+400}_{-300} & 13000^{+600}_{-600} & 12700 \pm 200
    \enddata
\end{deluxetable*}
%\end{longrotatetable}

%\begin{longrotatetable}
\begin{deluxetable*}{RRRRRRRR}
    %\tabletypesize{\scriptsize}
    \tablecaption{%
        Same as Table~\ref{tab:tspec_full}, but for the wavelength bin scheme centered
        around each species investigated for Figure~\ref{fig:tspec_species}. The resulting
        weighted mean white"-light curve depth is \MeanWLCDepthErrSp.
        \label{tab:tspec_species}
    }
    \tablehead{
        \colhead{Wavelength (\angstrom)} &
        \colhead{Transit 1} &
        \colhead{Transit 2} &
        \colhead{Transit 3} &
        \colhead{Transit 4} &
        \colhead{Transit 5} &
        \colhead{Combined}
    }
    \startdata
    5780.4 - 5825.4 & 13800^{+1600}_{-1600} & 11800^{+1300}_{-1300} & 12700^{+700}_{-700}   & 14000^{+400}_{-400} & 13800^{+800}_{-800}   & 13600 \pm 300 \\
5825.4 - 5870.4 & 14100^{+1500}_{-1600} & 12500^{+700}_{-1000}  & 13000^{+700}_{-700}   & 13600^{+400}_{-400} & 14300^{+900}_{-1000}  & 13500 \pm 300 \\
5870.4 - 5915.4 & 12400^{+1400}_{-1400} & 12200^{+1400}_{-1400} & 13000^{+1000}_{-1200} & 13300^{+500}_{-600} & 13800^{+800}_{-800}   & 13300 \pm 400 \\
5915.4 - 5960.4 & 13100^{+1400}_{-1400} & 12900^{+1300}_{-1300} & 13500^{+1100}_{-1100} & 14000^{+400}_{-500} & 13900^{+600}_{-600}   & 13800 \pm 300 \\
5960.4 - 6005.4 & 13300^{+1700}_{-1700} & 12000^{+700}_{-700}   & 12300^{+800}_{-900}   & 13900^{+300}_{-300} & 13800^{+700}_{-900}   & 13500 \pm 300 \\
7657.0 - 7667.0 & 12500^{+1500}_{-1400} & 12100^{+1700}_{-1600} & 12600^{+1800}_{-1800} & 12800^{+900}_{-800} & 12800^{+1100}_{-1000} & 12600 \pm 500 \\
7667.0 - 7677.0 & 12500^{+1500}_{-1400} & 11200^{+1100}_{-1100} & 13700^{+1400}_{-1500} & 12500^{+600}_{-700} & 13000^{+800}_{-800}   & 12500 \pm 400 \\
7677.0 - 7687.0 & 12700^{+1400}_{-1300} & 11700^{+1200}_{-1200} & 12300^{+1100}_{-1100} & 13100^{+700}_{-800} & 13700^{+800}_{-800}   & 12900 \pm 400 \\
7687.0 - 7697.0 & 11500^{+1000}_{-1100} & 12000^{+1400}_{-1300} & 10500^{+900}_{-900}   & 13800^{+700}_{-700} & 12900^{+700}_{-700}   & 12400 \pm 400 \\
7697.0 - 7707.0 & 12100^{+1400}_{-1400} & 11200^{+900}_{-900}   & 12800^{+1100}_{-1100} & 14200^{+600}_{-700} & 14500^{+700}_{-700}   & 13500 \pm 400 \\
8089.0 - 8129.0 & 12700^{+800}_{-800}   & 12100^{+900}_{-900}   & 11600^{+500}_{-500}   & 13800^{+700}_{-700} & 14200^{+500}_{-600}   & 12900 \pm 300 \\
8129.0 - 8169.0 & 12500^{+600}_{-700}   & 12200^{+1300}_{-1200} & 11500^{+500}_{-600}   & 14000^{+600}_{-600} & 14600^{+500}_{-500}   & 13200 \pm 300 \\
8169.0 - 8209.0 & 12700^{+1000}_{-900}  & 11800^{+1300}_{-1200} & 11800^{+700}_{-600}   & 13000^{+600}_{-600} & 14300^{+600}_{-600}   & 12900 \pm 300 \\
8209.0 - 8249.0 & 14100^{+1100}_{-1100} & 12600^{+1400}_{-1400} & 11700^{+600}_{-600}   & 13200^{+600}_{-700} & 14200^{+600}_{-700}   & 13100 \pm 300 \\
8249.0 - 8289.0 & 13200^{+900}_{-900}   & 11200^{+800}_{-900}   & 10800^{+500}_{-500}   & 13000^{+600}_{-600} & 14300^{+600}_{-700}   & 12300 \pm 300 \\
8669.0 - 8869.0 & 12600^{+700}_{-600}   & 12300^{+400}_{-500}   & 12800^{+400}_{-400}   & 12900^{+400}_{-400} & 12800^{+700}_{-700}   & 12700 \pm 200 \\
8869.0 - 9069.0 & 13100^{+800}_{-900}   & 12800^{+400}_{-500}   & 12300^{+600}_{-500}   & 12700^{+400}_{-300} & 13000^{+600}_{-600}   & 12700 \pm 200
    \enddata
\end{deluxetable*}
%\end{longrotatetable}

\begin{deluxetable*}{LRRRRRR}
    %\tabletypesize{\scriptsize}
    \caption{%
        Median and $1\sigma$ confidence levels of the retrieved parameters for the
        corresponding retrieved transmission spectra in Figure~\ref{fig:retrieval_tspec}.
    }
    \label{tab:retrieval_summary}
    \tablehead{\\
        \colhead{Parameter} &
        \colhead{\begin{tabular}[c]{@{}c@{}}TiO\\(clear)\end{tabular}} &
        \colhead{\begin{tabular}[c]{@{}c@{}}K\\(clear)\end{tabular}} &
        \colhead{\begin{tabular}[c]{@{}c@{}}Na+K\\(clear)\end{tabular}} &
        \colhead{\begin{tabular}[c]{@{}c@{}}Na+K\\(clear+cloud)\end{tabular}} &
        \colhead{\begin{tabular}[c]{@{}c@{}}Na+K\\(clear+haze)\end{tabular}} &
        \colhead{\begin{tabular}[c]{@{}c@{}}Na+K\\(clear+spot)\end{tabular}}\\
    }
    \startdata
    \log P_0                & 0.5^{+4}_{-4}        & 3.4^{+2}_{-3}       & 1.4^{+3}_{-3}        & 0.8^{+3}_{-3}        & 0.5^{+3}_{-3}        & -1.3^{+4}_{-3}        \\
    T                       & 1751.1^{+783}_{-804} & 620.1^{+328}_{-209} & 1938.0^{+652}_{-779} & 1827.7^{+714}_{-950} & 1878.4^{+709}_{-973} & 1393.6^{+1043}_{-886} \\
    \log\text{ TiO}         & -7.1^{+4}_{-4}       & --                  & --                   & --                   & --                   & --                    \\
    \log\text{ K}           & --                   & -2.6^{+2}_{-3}      & -19.4^{+8}_{-7}      & -18.7^{+8}_{-8}      & -18.8^{+9}_{-7}      & -19.0^{+8}_{-7}       \\
    \log\text{ Na}          & --                   & --                  & -4.0^{+3}_{-3}       & -3.9^{+3}_{-4}       & -4.8^{+3}_{-10}      & -15.5^{+9}_{-10}      \\
    \log\sigma_\text{cloud} & --                   & --                  & --                   & -37.7^{+10}_{-8}     & --                   & --                    \\
    \log a                  & --                   & --                  & --                   & --                   & -11.0^{+14}_{-13}    & --                    \\
    \gamma                  & --                   & --                  & --                   & --                   & -2.3^{+2}_{-1}       & --                    \\
    T_\text{star}           & --                   & --                  & --                   & --                   & --                   & 5906.8^{+349}_{-327}  \\
    T_\text{spot}           & --                   & --                  & --                   & --                   & --                   & 2200.0^{+0}_{-0}      \\
    f_\text{spot}           & --                   & --                  & --                   & --                   & --                   & 0.0^{+0}_{-0}
    \enddata
\end{deluxetable*}

\begin{figure*}[htbp]
    \centering
    \includegraphics[width=0.98\linewidth]{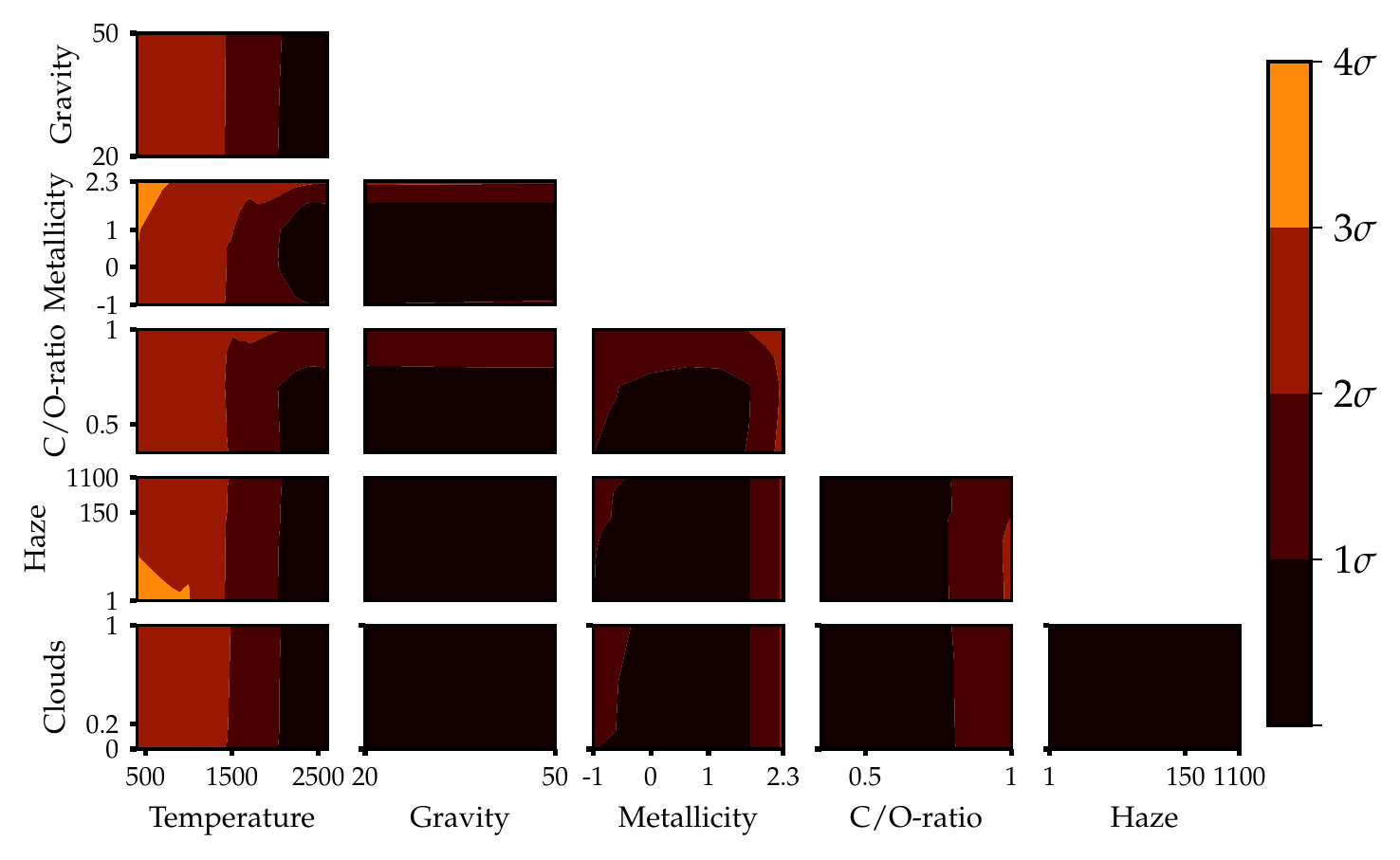}
    \caption{%
        $\chi^2$ map associated with Figure~\ref{fig:tspec_forward_model}, showing all
        combinations of the above grid parameters explored in the local condensation
        \texttt{ATMO} generic grid. The colorbar on the right indicates the confidence
        interval of each contour.
        \pylink{https://icweaver.github.io/HAT-P-23b/notebooks/tspec_forward_model_corner.html}
    }
    \label{fig:tspec_forward_model_corner}
\end{figure*}

\end{document}